\documentclass[]{raa}            % referee version: for submission

\usepackage{graphicx,times}             %for PS/EPS graphics inclusion, new
\usepackage{natbib}
\usepackage{amssymb,amsmath}
\usepackage{booktabs}
\usepackage{capt-of}
\bibpunct{(}{)}{;}{a}{}{,}

\usepackage[a4paper=true,pagebackref=true]{hyperref}
\hypersetup{colorlinks = true, linkcolor = green, anchorcolor = red, citecolor = blue, filecolor = red, pagecolor = red, urlcolor = red}

\begin{document}

   \title{Absolute Parameters of Young Stars: PU Pup}

   \volnopage{Vol.0 (20xx) No.0, 000--000}      %%preserved for Editor. Don't remove!
   \setcounter{page}{1}          %%starting page, preserved for Editor. Don't remove!

   \author{A. Erdem
      \inst{1,2}
   \and D. S\"{u}rgit
      \inst{1,3}
   \and Timothy S. Banks
      \inst{4,5}
    \and B. \"{O}zkarde\c{s}
        \inst{1,3}
    \and E. Budding
        \inst{6,7,8,9}
   }

   \institute{\c{C}anakkale Onsekiz Mart University, Astrophysics Research Center and Ulupınar Observatory, TR-17100, Çanakkale, Turkey\\
        \and
             Department of Physics, Faculty of Arts and Sciences, \c{C}anakkale Onsekiz Mart University, Terzio\u{g}lu Kamp\"{u}s\"{u}, TR-17100, \c{C}anakkale, Turkey \\
        \and
             Department of Space Sciences and Technologies, Faculty of Arts and Sciences, \c{C}anakkale Onsekiz Mart University, Terzio\u{g}lu  Kamp\"{u}s\"{u}, TR-17100, \c{C}anakkale, Turkey\\
        \and
            Data Science, Nielsen, 200 W Jackson, Chicago, IL 60606, USA; {\it tim.banks@nielsen.com}\\
        \and
            Physics \& Astronomy, Harper College, 1200 W Algonquin Rd, Palatine, IL 60067, USA\\
        \and
            Visiting astronomer, Mt.\ John Observatory, University of Canterbury, Private Bag 4800, Christchurch 8140, NZ\\
        \and
            Carter Observatory, Wellington 6012, New Zealand\\
        \and    
            School of Chemical \& Physical Sciences, Victoria University of Wellington, Wellington 6012, New Zealand\\
        \and    
            Variable Stars South, RASNZ, PO Box 3181, Wellington, New Zealand\\
\vs\no
   {\small Received~2021~March~19; accepted~2021~July~09}}

%------------------------------------------------------------------------------------------

\abstract{ We present combined photometric and spectroscopic analyses of the
southern binary star PU Pup. High-resolution spectra of this system were
taken at the University of Canterbury Mt. John Observatory in the years
2008 and again in 2014-15. We find the light contribution of the
secondary component to be only $\sim$2\% of the total light of the
system in optical wavelengths, resulting in a single-lined spectroscopic
binary. Recent TESS data revealed grazing eclipses within the light
minima, though  the tidal distortion, examined also from  HIPPARCOS
data, remains the predominating light curve effect. Our model shows PU
Pup to have the more massive primary relatively close to filling its
Roche lobe. PU Pup is thus approaching the rare `fast phase' of
interactive (Case B) evolution. Our adopted absolute parameters are as
follows: 
$M_1$ =  4.10 ($\pm$0.20) M$_{\odot}$, $M_2$ = 0.65 ($\pm$0.05) M$_{\odot}$,
$R_{1}$ = 6.60 ($\pm$0.30) R$_{\odot}$, $R_2$ = 0.90 ($\pm$0.10) R$_{\odot}$;
$T_{1}$ = 11500 ($\pm$500) K, $T_{2}$ = 5000 ($\pm$350) K;
photometric distance = 186 ($\pm$20) pc, age = 170 ($\pm$20) My. The
less-massive secondary component is found to be significantly oversized
and overluminous compared to standard Main Sequence models. We discuss
this discrepancy referring to heating from the reflection effect.
}

\authorrunning{Erdem~{\em et al.}}            %author_head in even pages
\titlerunning{Absolute Parameters of Young Stars: PU Pup}  % title_head in odd pages

\maketitle

%% The author head (on even pages) and the title head (on odd pages) will be
%% automatically extracted from \author{} and \title{}. Whenever the title is too long,
%% you will be asked to supply a shorter one by inserting either \authorrunning{} or
%% \titlerunning{} before \maketitle. Anyway, you can specify your own heads.
%%
%%
%% Note: In the following text body of your manuscript, please note several differences from
%%       other major journals:
%% (1) \subsection{Please Capitalize the First Letter of Each Notional Word in Subsection Title}
%% (2) Please Capitalize the First Letter of Each Notional Word in all tables' captions

%------------------------------------------------------------------------------------------

\section{Introduction}
\label{sec1}  

The system PU Puppis (= HR2944, HD61429, HIP 37173), also known as m Pup,
is a relatively bright (B = 4.59, V = 4.70) early type giant
\citep[B8III,][]{garrison94} located at a distance of about 190 pc with the
galactic co-ordinates $\lambda \sim$240\fdg 7, $\beta \sim$ --1\fdg 8.
\citet{garrison94} noted its strong rotation, though without any
significant effect apparent in its 4-colour photometric indices. While
\citet{jaschek69} have given the spectral type of PU Pup to be B9V from
its $U B V$ colour indices, \citet{stock02} determined the type to be
B8IV from the system's ($B-V)_0$.

%Testing the editing commands supplied by the journal.  Works!
%blabla...  

The photometric variability was announced by \citet{stift79} during a
programme monitoring other stars in the sky region nearby.   Stift
noticed the variable had been reported as a component of the close
visual double ADS 6246, the companion being of similar spectral type and
magnitude, with a separation of about 0.1\arcsec\ or about 20 AU
perpendicular to the line of sight. A corresponding period of the order
of 30 years or greater for this wide pair might then be expected. Stift
surmised that the close binary might be of the W UMa or $\beta$ Lyr
kinds, but the relatively early type and long period would make PU Pup a
very atypical representative of the first of these classes. The star appears to
have received relatively little individual attention, but its
variability was confirmed by the HIPPARCOS satellite \citep{esa}.

\citet{vanden27} first reported that m Pup was a close visual double and
it appears as number 731 in his fourth list of new southern doubles,
with the remark ``too close''. The Washington Visual Double Star
Catalogue 1996.0 \citep{worley97} added the note ``Duplicity still not
certain.  Needs speckle.''  The latest (internet) edition of the
Washington Double Star Catalogue mentions that the double star
identification now seems likely to be spurious, as the system was
unresolved by the high altitude SOAR telescope on 5 occasions between
2009 and 2013.

\citet{stift79} gave the period of the variable to be 2.57895 d with the
primary minimum epoch 2443100. This is significantly different from the
period given by HIPPARCOS of 2.58232 days, at epoch 2448501.647, although the
form of the light curve given by Stift appears essentially similar to
that of HIPPARCOS. \citet{stift79} considered other period
possibilities, suggesting uncertainty in his evaluation, nevertheless
the original period was retained in recent editions of the GCVS
\citep{samus17}. The Eclipsing Variables catalogue of
\citet{svechnikov90} lists PU Pup as having a $\beta$ Lyr type light
curve, composed of two near Main Sequence stars with the orbital
inclination of about 70.5 deg, the relatively shallow minima being
presumably offset by the light of the supposed third component, which
would contribute about 45\% of the photometrically monitored light,
using the magnitude difference of \citet{vanden27}. The two minima, at
around 0.05 and 0.04 mag depth, appeared too shallow for the system to
be regarded as eclipsing without third light; however, an ellipsoidal
type variability may still fit the data if there is little or no third
light. 
\citet{budding19} solved the HIPPARCOS light curve of PU Pup and proposed a relatively close
binary star model with a rather low mass ratio and low inclination to give the ellipsoidal form of its
light variation.

Recent high accuracy TESS monitoring \citep{ricker14} of PU Pup showed
that, although the predominating effect in the light curve is that of
the tidal distortion (`ellipticity effect'), grazing eclipses do occur at
the base of the two minima per orbit, allowing a firm constraint on the orbital
inclination.

%------------------------------------------------------------------------------------------

\begin{figure}
  \centering
	\includegraphics[scale=0.48]{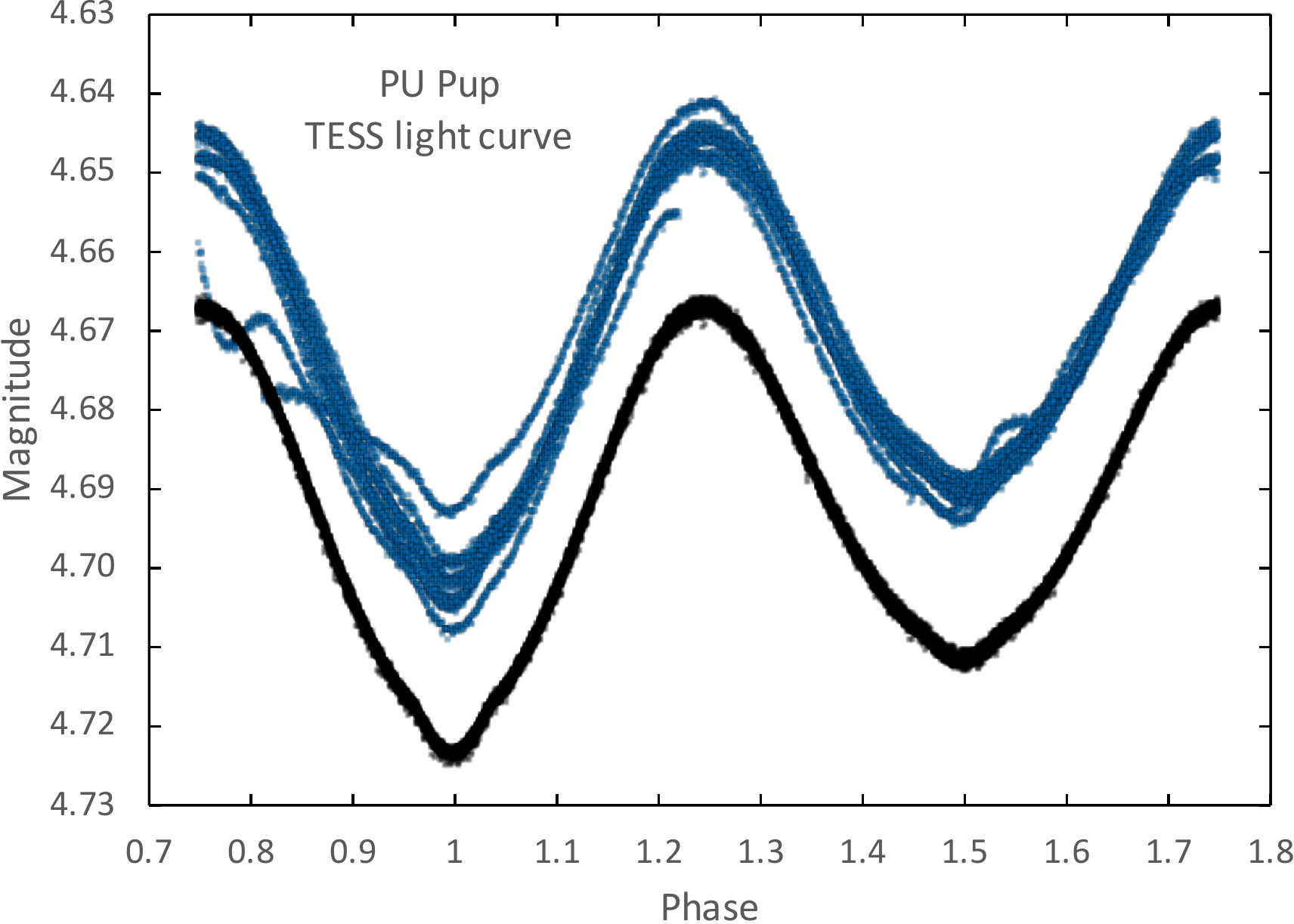}
\caption{TESS light curves of PU Pup. The lower (black) points show the SAP data, 
while the upper (blue) points display the PDCSAP data from the MAST archive. The
vertical axis gives the magnitude of the system, while the horizontal axis is the orbital
phase as per equation~\ref{eqn1}}
\label{tess}
\end{figure}

\section{Orbital Period}
\label{sec2}

\begin{figure}
  \centering
	\includegraphics[scale=0.70]{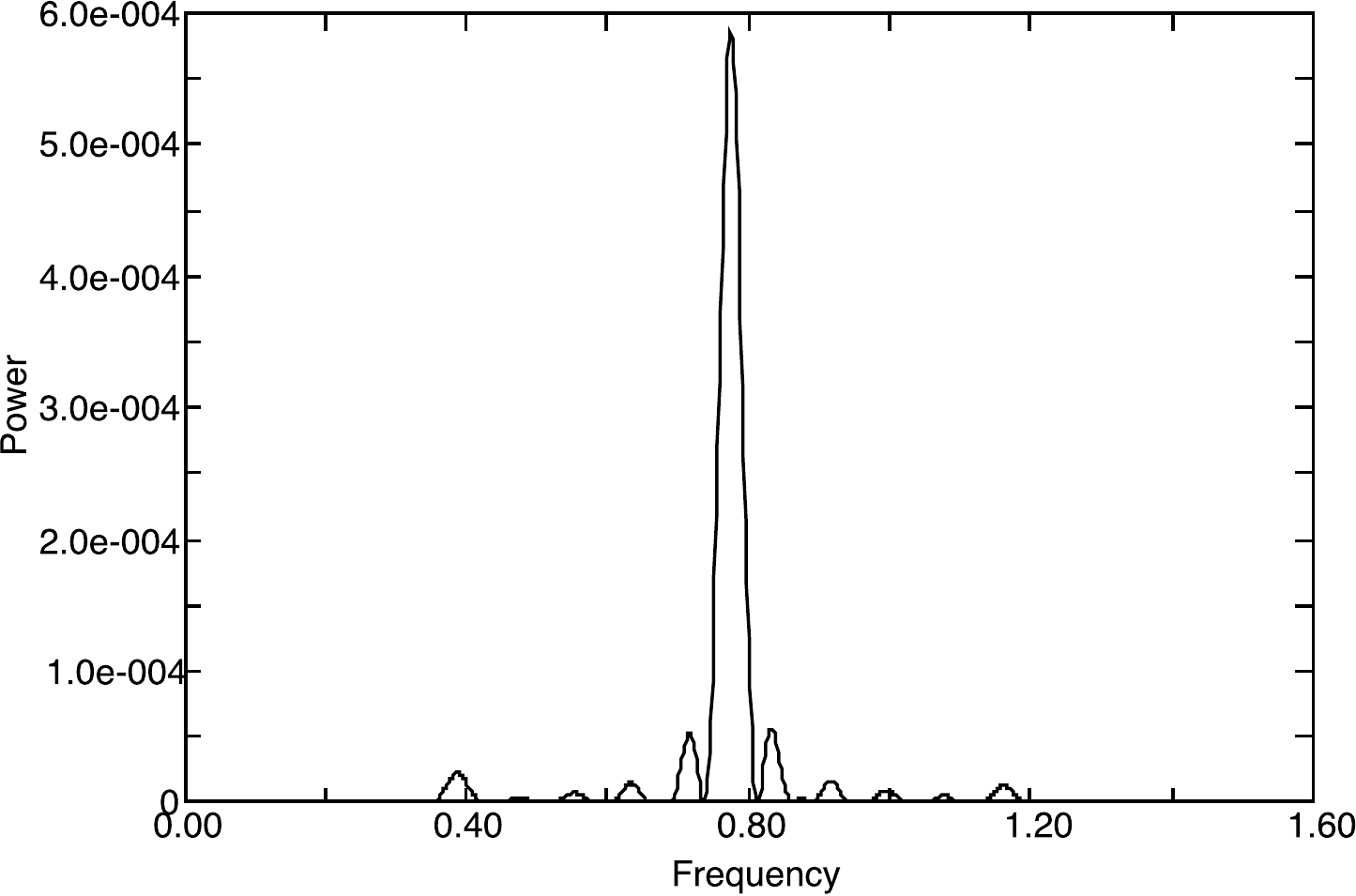}
\caption{The Fourier spectrum of TESS data of PU Pup, plotting frequency on the horizontal
axis and power on the vertical axis.}
\label{fourier}
\end{figure}

Previous differences in the period values, mentioned in Section
\ref{sec1}, prompted us to check this parameter. 
PU Pup (TIC 110606015) was observed by Transiting Exoplanet Survey
Satellite (TESS)  during Sector 7 of the mission, with two-minute cadencing. These
light curves  were downloaded from the Mikulski Archive for Space
Telescopes (MAST) \citep[cf.][]{jenkins16}. We used the straight Simple
Aperture Photometry (SAP) data, since the Pre-search Data Conditioning
Simple Aperture Photometry (PDCSAP) is optimized for planet transits and
it appears that the PDCSAP detrending leads to additional effects in the
13.7 d period light curves of PU Pup (see Figure \ref{tess}). After
outliers (having non-zero `quality' flags) were removed from the SAP data,
about 17,000 points remained for analysis. With its high inherent
accuracy (each datum is of the order of 0.0001 mag) the TESS light curves
of PU Pup allowed us to discern small eclipse effects at the base of
both light minima (see Figure \ref{tesslc}). These 
effects would not have been noticeable in typical Earth-based
photometry, or even that of HIPPARCOS. Fourier analysis of
TESS light curves of the system was carried out using the program {\sc
Period04} \citep{lenz05}. 
As can be seen from Figure \ref{fourier}, only
one dominant frequency was obtained as $f_1 = 0.77448 \: (\pm \: 0.00003)$
cycle/days. The period equivalent of this frequency is 1.29119 days, which
corresponds to  half of the orbital period; i.e.\ $P_{orb} = 2.58237 \: (\pm  \: 0.00005)$ d.

The TESS satellite observed 10 primary minima and 8 secondary minima of
PU Pup over two observation cycles. The flux measurements were converted
to magnitudes, and times of these minimum light were computed using a
standard {\sc fortran} program based on the K-W method (\citealt{kwee56}; see also 
\citealt{ghedini82}). These times of minima are given
in Table \ref{tableA1} of the Appendix.

\begin{figure}
  \centering
	\includegraphics[scale=0.60]{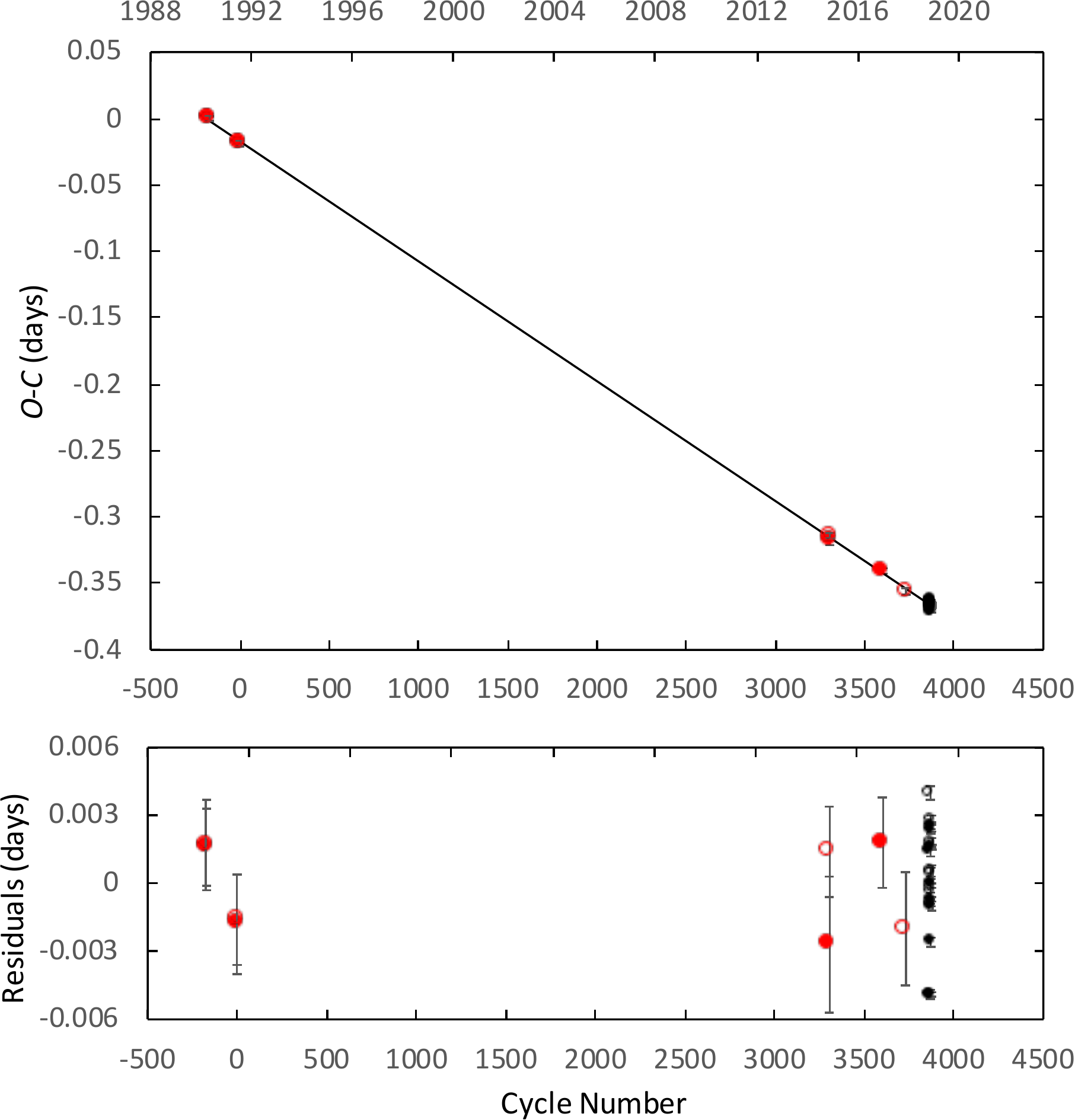}
\caption{$O-C$ diagram of PU Pup: the $O-C$ values for times of
minima (obtained directly from the TESS observations,
are shown as black circles) while the $O-C$ values for times of
minima (derived using the theoretical LC templates) are given as
red circles. The $O-C$ values for the times of primary
and secondary minima are also indicated by filled and hollow
symbols respectively. The bottom panel shows the residuals from the 
linear model. Cycle number is
the number of orbits from the starting epoch given in equation~\ref{eqn1}.}
\label{o-c}
\end{figure}

We sought to check on $O - C$ variations for a change in the orbital
period.  Although PU Pup is a relatively bright star, since the
minimum depths of its light curves are very shallow, literature  times
of minima are few and sparsely arranged in time. 
We found only the KWS
$V$ and $I$ data \citep[from the Kamogata/Kiso/Kyoto Wide-Field
Survey;][]{maehara14}, apart from  those of \citet{stift79}, HIPPARCOS
and TESS mentioned already. 
We applied a method similar to the automated
fitting procedure of \citet{zasche14} to derive individual times of
minima of PU Pup from HIPPARCOS and KWS. The $O - C$ (observed minus
calculated times of minima) values, listed in the Appendix, were
examined from such findings. 
Our are plotted against the cycle number
and observation years in Figure \ref{o-c}.
The $O-C$ diagram of PU Pup shows a linear trend.
Using a linear ephemeris model and
least-squares optimization, the following result was obtained:
\begin{equation}
\label{eqn1}
	T_{min} = 2448501.6306(11) + 2.5822297(3) \times E  .
\end{equation}

Comparing the mean period values from the epoch of \citet{stift79} to
the latest TESS times of minima (Table \ref{table3}) there is a
suggestion that the period is shortening. A typical estimate of $\Delta
P/P$ is $\sim  -3 \times 10^{-8}$  per orbit or $ -4\times 10^{-6}$ per year, over the
last $\sim 40$ years. But the main result is that while we may derive an
improved mean period and reference epoch, the data are too sparse for
reliable period analysis. 
Although the $O-C$ diagram in Figure \ref{o-c} does not give 
any clues to support the third light estimated from the light curve modeling (Section \ref{sec4}) or any changes in the orbital period of PU Pup, 
it is also noted that the scatter in the TESS data is larger than the 
expected timing accuracy ($\sim$ 5 minutes) 
suggestive of perhaps some short term irregularity.

\begin{table}
\caption{Mean Periods Over Longer Time Intervals.  The Hipparcos
epoch and period is used for reference. `TOM' stands for Time of Minimum, $P$ the orbital
period, and $\Delta P/P$ the linear change in period per orbit.
\label{table3}} 
%\scriptsize 
\begin{center}
\begin{tabular}{lllll} 
\hline
\multicolumn{1}{c}{Epoch} 
&\multicolumn{1}{c}{ToM}& \multicolumn{1}{c}{Orbits} & 
\multicolumn{1}{c}{Mean $P$} &\multicolumn{1}{c}{$\Delta P/P$} \\
\multicolumn{1}{c}{ HJD240000+} 
&\multicolumn{1}{c}{}& \multicolumn{1}{c}{} & 
\multicolumn{1}{c}{} &\multicolumn{1}{c}{} \\ 
\hline
43099.4336 & 43100.00	  & -2092 & 2.5820492 &  $-1.29\times 10^{-7}$ \\
54805.0901 & 54804.8319  & 2441  & 2.5822142 &  $-4.33\times 10^{-8}$ \\
56669.5252 & 56669.2096  & 3163  & 2.5822202 &  $-3.15 \times 10^{-8}$  \\
\hline
\end{tabular}
\end{center}
\end{table}

%------------------------------------------------------------------------------------------

\section{Spectroscopy}
\label{sec3}

Spectroscopic data on PU Pup have been collected using the High
Efficiency and Resolution Canterbury University Large \'{E}chelle
Spectrograph (HERCULES) of the Department of Physics and Astronomy,
University of Canterbury, New Zealand \citep{hearnshaw02} on the 1 m
McLellan telescope of the Mt John University Observatory (MJUO) near
Lake Tekapo ($\sim$43\degr 59' S, $\sim$174\degr 27' E).  A few dozen
high-dispersion spectra have been taken of the system over the last
decade. A reasonable phase coverage of the system's radial velocity
variations was first collected in 2008. 

To attain a fair signal to noise (S/N) ratio (typically around 100) the
100 micron optical diameter fibre cable has generally been used.  This
enables a theoretical resolution of $\sim 40,000$.   Typical exposure
times were about 500 seconds for the SI600s CCD camera.  The raw data were
reduced with {\sc HRSP} 5.0 and 7.0 \citep{skuljan14,skuljan20}, the
\'{e}chelle's spectral orders between 85 and 125 being convenient for
stellar spectral image work.   We applied {\sc IRAF} tools to the {\sc
HRSP}-produced image files in order to determine  information
such as the radial and rotational velocities. The {\sc IRAF} routine {\sc
SPLOT} was useful in this context.

Spectral orders and lines used for line identifications and RV measurements are given in Table \ref{lines}.
The plots of spectral orders selected from the spectrum of PU Pup taken on the night of December 03, 2015 and 
at the conjunction phase are also shown in Figure B-1  in Appendix \ref{appendix_b} as an example.
Non-hydrogen lines that could be measured well (Table \ref{lines}: the
He I lines and the Si II $\lambda$5056 feature) have a depth of
typically $\sim$3\% of the continuum. Such lines are relatively wide,
with a width at base of typically $\sim$6 {\AA} or $\sim$220 pixels. 
The hydrogen lines are, of course, much better defined, but they are
very broadened and perhaps complicated by dynamical effects in the
system or over component surfaces. The positioning of a well-formed
symmetrical (non-H) line can be expected to be typically achieved to
within $\sim$10 pixels or (equivalently) up to $\sim$10 km/s.   The
measured lines move in accordance with only one spectrum, the system is
thus `single lined' and no feature unequivocally from the secondary
could be definitely  identified.

The equivalent width ({\sc ew}) of H$\beta$ was measured to be typically
$\sim$7.5, and that of He~I 6678 to be $\sim$0.12.  While the He~I {\sc ew} points to a
spectral type of about B8, the corresponding {\sc ew} value for H$\beta$ would
be higher for a normal dwarf at B8.  The issue is resolved by the lower
gravity  luminosity classifications of \citet{garrison94} and
\citet{stock02}.  

\begin{table}
\caption{Spectral Line Identifications.
Asterisks refer to the appearance of the same line on successive  orders.
\label{lines}}
\begin{center}
\begin{tabular}{lcll}
\hline
\multicolumn{1}{l}{Species}  & \multicolumn{1}{c}{Order no.} &
\multicolumn{1}{c}{$\lambda$} & \multicolumn{1}{l}{Comment}   \\
\hline
He I & 85        &  6678.149    &  strong   \\
H$\alpha$ & 87   &  6562.81     &  not used for rv  \\
Si II    &  89   &  6371.359    &  moderate    \\
Si II    &  90   &  6347.10     &  moderate    \\
He I &  97       &  5875.65     &  strong  \\
Fe II & 107      &  5316.609    &  moderate    \\
Fe II & 110       &  5169.03    &  moderate     \\
Si II  & 112      &  5056.18    &   strong$^{*}$     \\
Si II  & 113      &  5056.18    &   strong$^{*}$     \\
H$\beta$ & 117   &  4861.3      &  not used for rv \\
Fe II & 124      &  4583.829    &  weak \\
Fe II & 125       &  4549.50    & Fe II + Ti II blend  \\
\hline
\end{tabular}
\end{center}
%\begin{tablenotes}
%\item \textit{Notes:} $^{*}$ Orders 112 and 113 overlap in this region.
%\end{tablenotes}
%Orders 112 and 113 overlap in this region.
\end{table}
 
\subsection{Radial Velocities}
\label{sec3.1}

 Mean wavelengths were derived using the IRAF SPLOT tool (k-k) on the
strong lines in Table \ref{lines} as taken from MJUO observations made in
2008, 2014 and 2015. These produced the representative radial velocities (RVs) 
given in Table \ref{RVs}. The listed dates and velocities  have
been corrected to heliocentric values with the aid of the HRSP and IRAF
program suites.

 \begin{table}
\begin{center}
\caption{Radial Velocity Data for PU Pup.  Error estimates, typically 
of order 5 km s$^{-1}$ for individual measures, are indicated in
Figure \ref{fig_rv}, and discussed in the text.  
\label{RVs}} 
   \begin{tabular}{cccc|cccc} 
\hline
\multicolumn{1}{c}{No}& \multicolumn{1}{c}{HJD}   & \multicolumn{1}{c}{Phase} &
\multicolumn{1}{c|}{RV}    &
\multicolumn{1}{c}{No}& \multicolumn{1}{c}{HJD}   & \multicolumn{1}{c}{Phase} &
\multicolumn{1}{c}{RV}    \\
\multicolumn{1}{c}{} & \multicolumn{1}{c}{245+}  &\multicolumn{1}{c}{} &
\multicolumn{1}{c|}{km s$^{-1}$}  &
\multicolumn{1}{c}{} & \multicolumn{1}{c}{245+}  &\multicolumn{1}{c}{} &
\multicolumn{1}{c}{km s$^{-1}$}  \\
\hline \\
1   	&	4802.9063	& 0.154 	& -9.0  &	 1	&	6667.0814	& 0.053 &  1.4  \\ 
2	&	4803.0954	& 0.228	&  1.9  &	 2		&	6668.0849	& 0.442	& 41.1  \\
3	&	4803.1365	& 0.243	&  7.5 &	3	&	6668.8805	& 0.750	& 52.3  	\\	
4	&	4803.9004	& 0.539	& 47.4  &	4	&	6668.9080	& 0.761	& 52.9  \\
5	&	4803.9124	& 0.544	& 54.3  & 5	&	6668.9366	& 0.772	& 52.0  \\
6	&	4803.9554	& 0.560	& 56.2  & 6	&	6669.8787	& 0.137	&  1.4	\\
7	&	4804.9119		& 0.931	& 26.3 & 7		&	6669.8871	& 0.140	&  -0.5  \\	
8	&	4804.9812	& 0.958	& 14.8  & 8	&	6669.9238	& 0.154	&  0.1  \\
9	&	4805.0601	& 0.988	& 13.1  &	9	&	6670.0036	& 0.185	&  0.9  \\
10 	&	4805.0946	& 0.002	& 22.3  & 	10	&	6670.8790	& 0.524	& 52.0  \\
11	&	4805.1247	& 0.013	&  7.7  & 11	&	6670.9174	& 0.539	& 57.2  \\
12  	&   	4805.1478	& 0.022	&  6.0  & 12	&	6671.0332	& 0.584	& 59.8  \\
13	&	4806.9436	& 0.718	& 56.9  & 13	&	6671.8804	& 0.912 	& 23.2	\\	
14	&	4806.9594	& 0.724	& 58.0 & 14	&	6674.8775	& 0.073	&  2.1  \\	
15	&	4807.0531	& 0.760	& 53.8  & 15	&	6675.9159	& 0.475	&  46.5  \\
16	&	4807.0595 	& 0.762	& 58.3  & 16	&	6675.9567	& 0.490	&  47.6  \\
\cmidrule(lr){5-8}
17	&	4807.1052	& 0.780	& 51.7  & 1   & 7354.0749   	& 0.091 	&  2.3  \\	
18	&	4807.1101		& 0.782	& 51.7  &	2   & 	7356.0368   	& 0.851	& 35.5	\\
19	&	4807.1157		& 0.784	& 52.0  &	3   &	7356.9445	& 0.202	&  5.6	\\
%\hline \\
%	&	&	&	&	1   & 7354.0749   	& 0.091 	&  2.3  \\	
%	&	&	&	&	2   & 	7356.0368   	& 0.851	& 35.5	\\
%	&	&	&	&	3   &	7356.9445	& 0.202	&  5.6	\\
	&	&	&	&	4   &	7357.9875	& 0.606	& 57.5	\\
	&	&	&	&	5   &	7360.0906	& 0.421	& 41.0	\\
	&	&	&	&	6   &	7360.9733	& 0.762	& 51.7	\\
\hline
\end{tabular}
\end{center}
\end{table}

\begin{figure}
  \centering
	\includegraphics[scale=0.53]{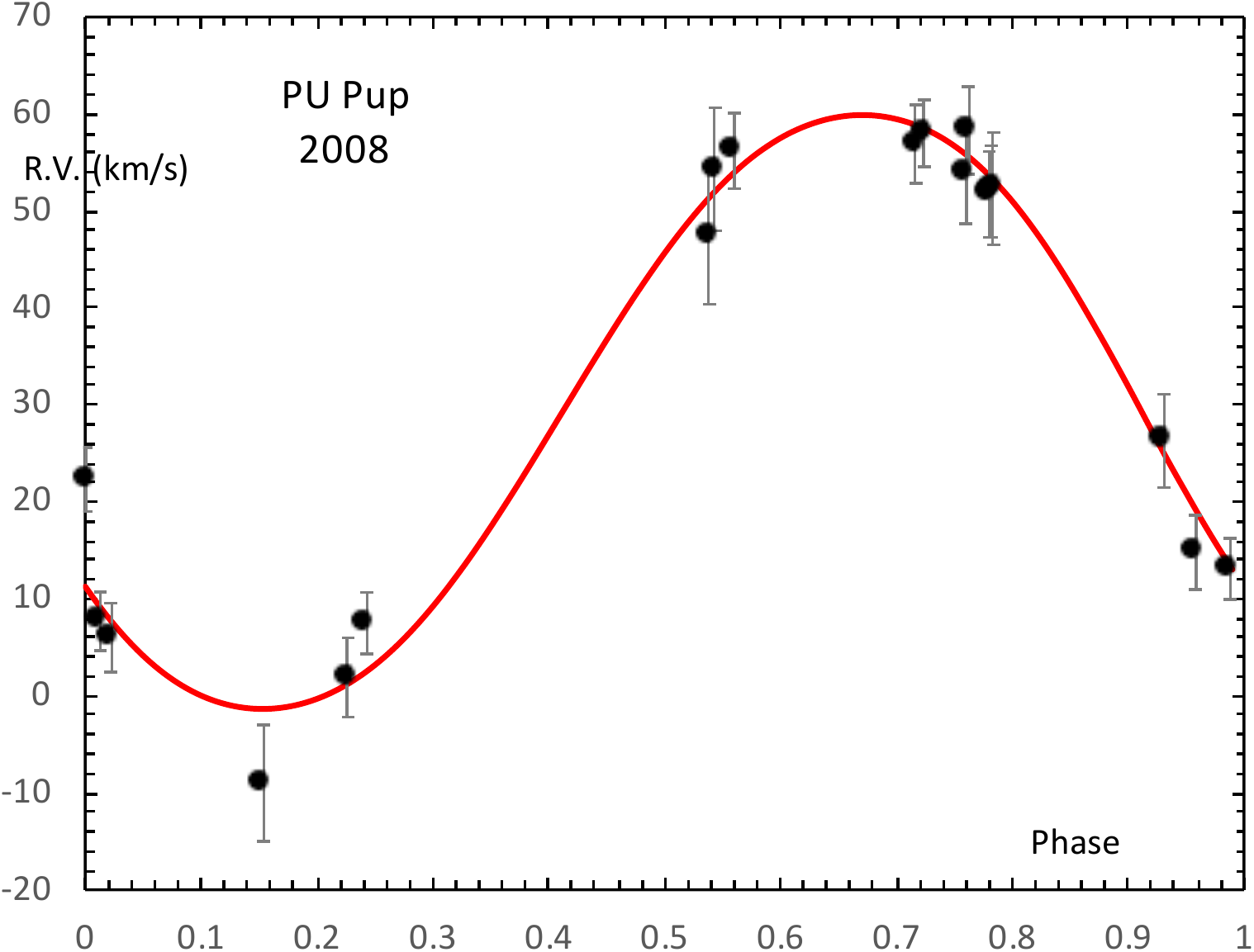} \\
	\includegraphics[scale=0.53]{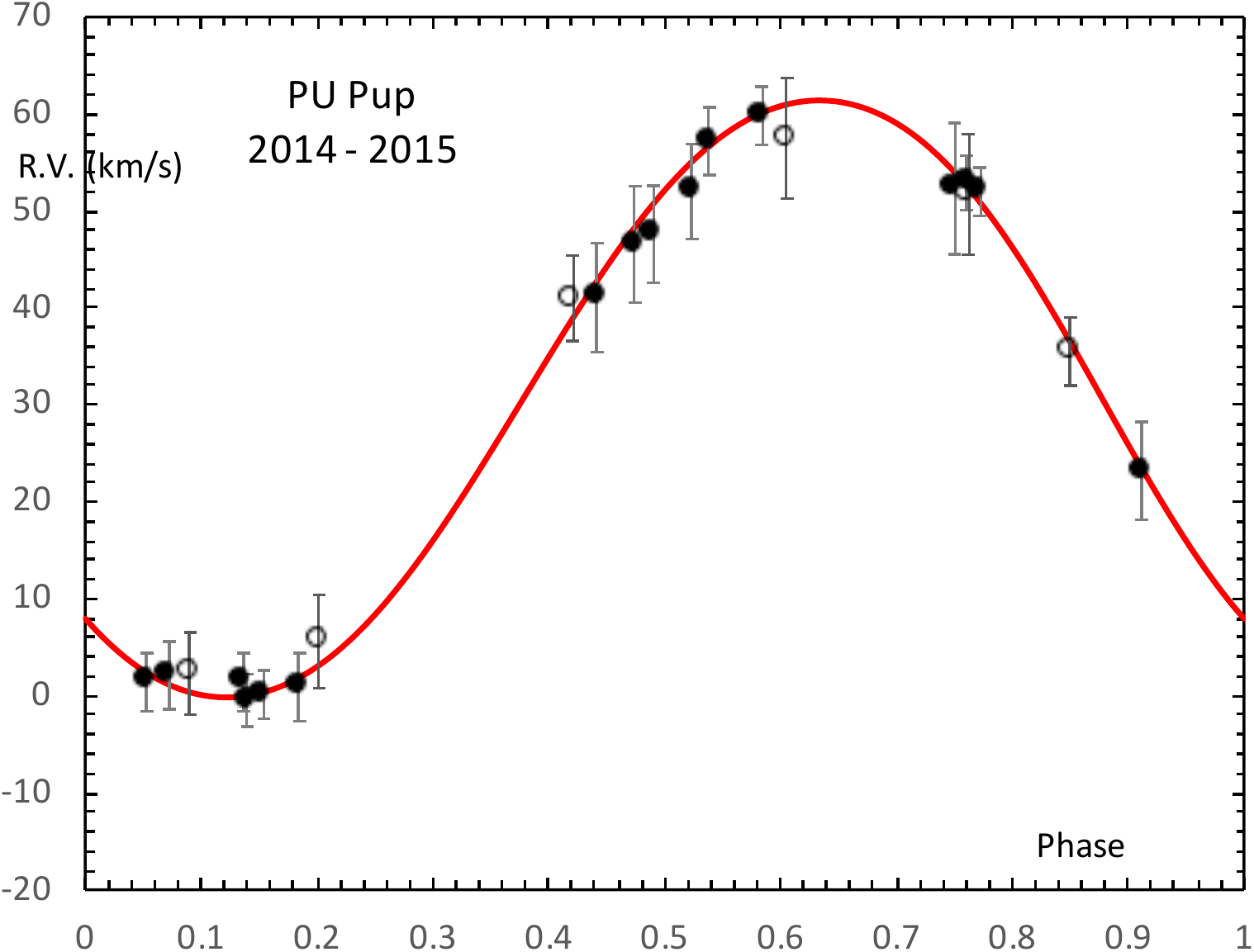}
\caption{Our best-fitting binary RV curves for PU Pup: 2008 data (upper panel), 
and 2014 \& 2015 data (lower panel).
Hollow circles show 2014 data, while empty circles display 2015 data.
The red lines represent the best fitting models. Orbital phase is
given as the horizontal axis, and radial velocities in km/s on the vertical axis.}
\label{fig_rv}
\end{figure}

The RVs from Table \ref{RVs} were fitted with an optimized binary system
model using the program {\sc fitrv4a}. The results are shown
in Figure \ref{fig_rv}, and the corresponding parameters listed in Table
\ref{RV_model}.

\begin{figure}
  \centering
	\includegraphics[scale=0.45]{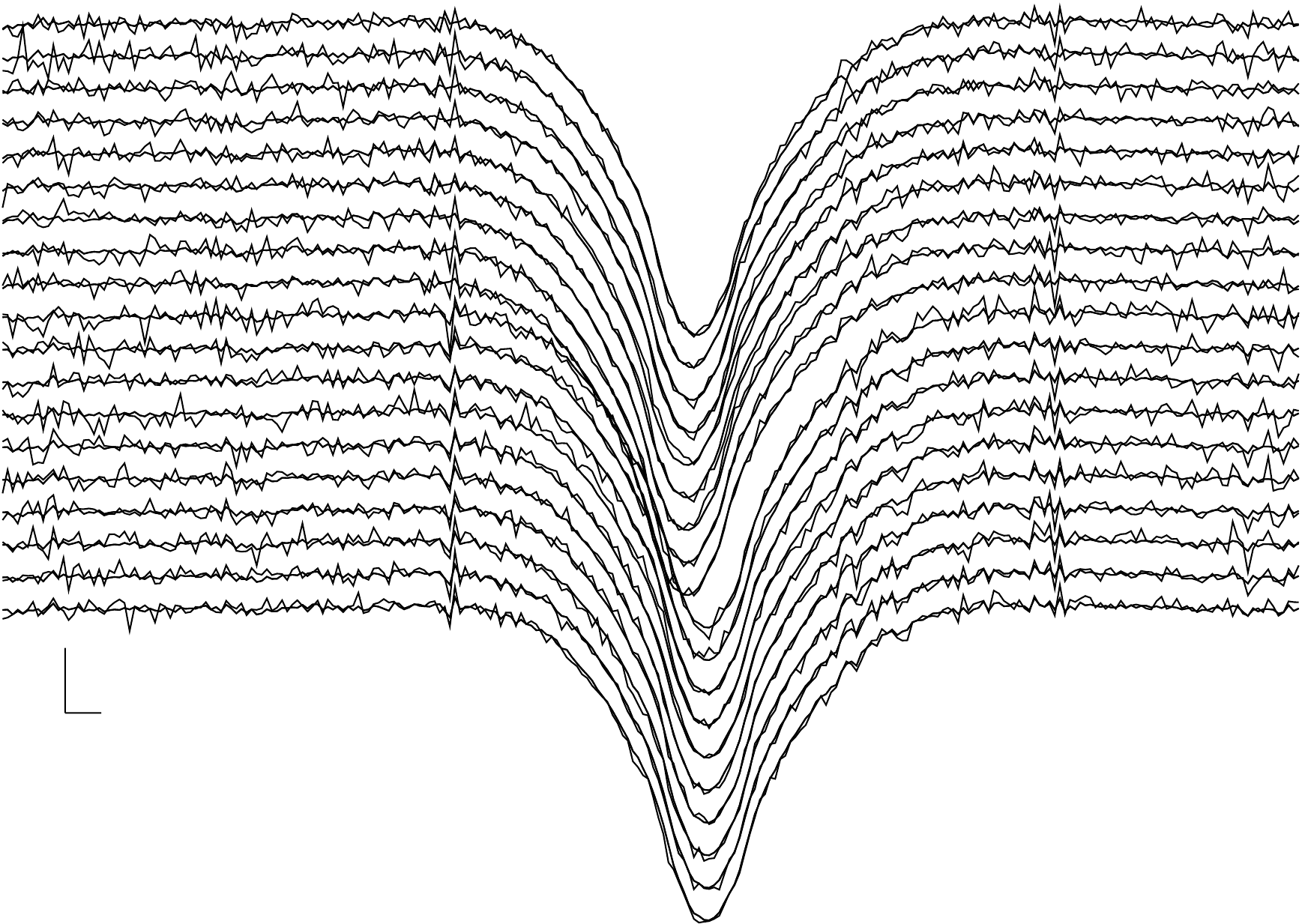} \\
	\includegraphics[scale=0.45]{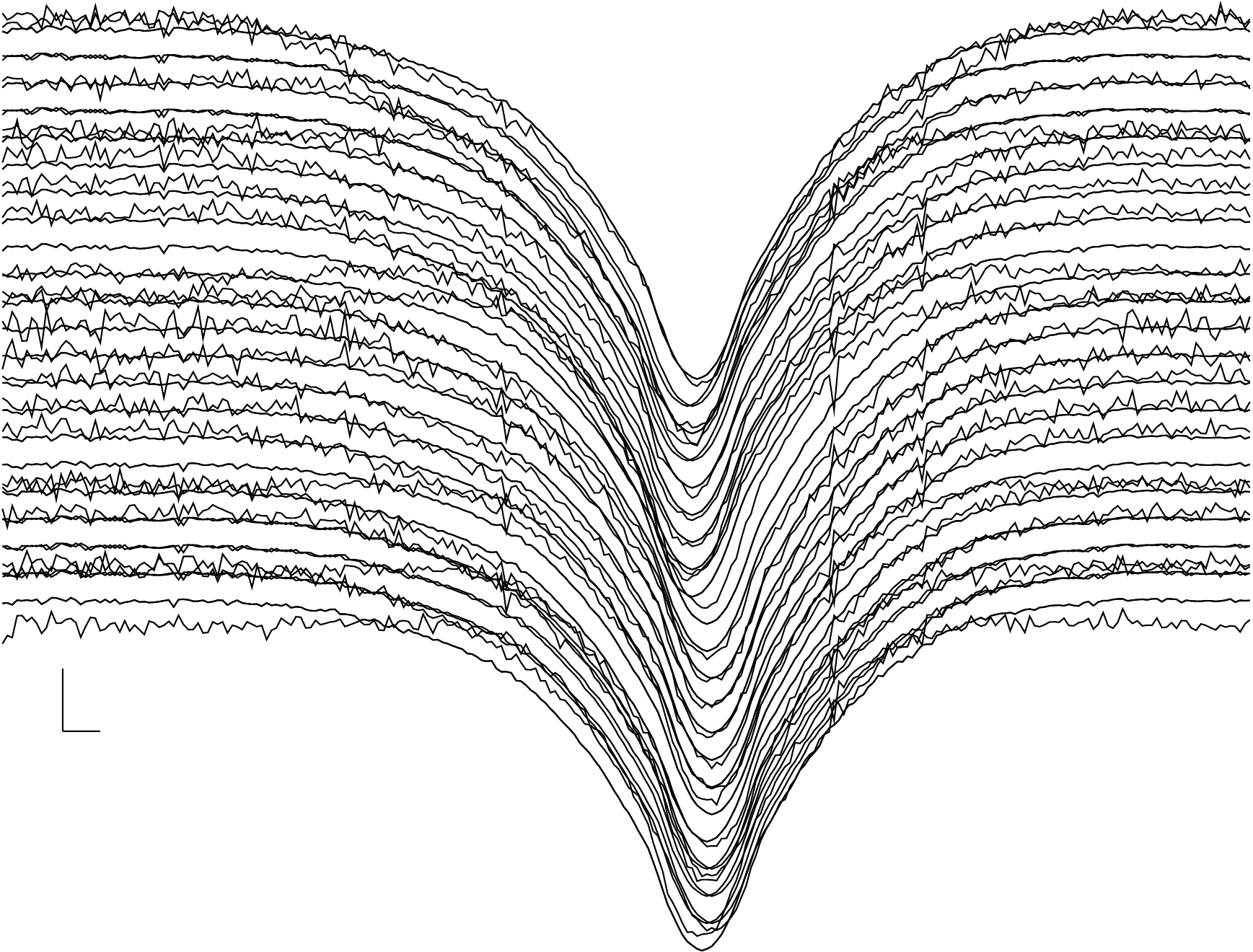}
\caption{{\sc korel} plot of H$\beta$ lines, from the 2008 data-set (upper panel),
and 2014 \& 2015 data-sets (lower panel) showing orbital motion.
}
\label{fig_korel}
\end{figure}

The {\sc korel} program \citep{hadrava04} was also used to check the RV
measurements in Table \ref{RVs} and spectroscopic orbital solutions
in Table \ref{RV_model} and to increase their reliability. For this, the
prominent H$\beta$ lines were chosen. The spectra of the secondary
component were not disentangled, it being assumed that -- in this
particular case of such a relatively low luminosity companion --
H$\beta$ is effectively unblended. The {\sc korel} fits and line
profiles of the H$\beta$ line are shown in Figure \ref{fig_korel}.

The mean epoch of the 2008 RV curve is HJD 2454805.0901. We may note
from Figure \ref{fig_rv} (upper panel) that the epoch of zero phase on
the rv curve has occurred before the ephemeris-predicted zero phase by
an interval of 0.0883. In other words, there has developed an `$O - C$'
of $-0.228$ days over the 2,441 orbital cycles after the HIPPARCOS zero
phase epoch. This implies a noticeable rate of secular period decrease
of order $\Delta P/P \sim 3\times10^{-8}$. If a similar calculation is
made for the 2014 --- 2015 radial velocity solutions as given in Table \ref{RV_model}, a
period decrease of the same order is found. 

% Should we realign the y axis labels in the following chart?

\begin{table}
\begin{center}
\caption{Binary Model RV Parameters for PU Pup. $K_1$ is the amplitude, $\phi_0$ the phase
offset, and $V_\gamma$ the mean system velocity in km/s. $f(M)$ is defined by equation~\ref{eqn2}.}
\label{RV_model}
\begin{tabular}{lcc}
\hline
Parameter		& 2008	& 2014 -- 2015 \\
\hline \\
$K_1$ (km/s)				& $31.5 (\pm1.3)$	& $31.4 (\pm0.5)$  \\
$\phi_0$ (deg)				& $-31.8 (\pm2.5)$	& $-44 (\pm1)$  \\
$V_\gamma$				& $29.3 (\pm1.0)$	& $30.6 (\pm0.4)$  \\
$a_{1}{\sin}i$ (R$_{\odot}$)	& 1.61 (${\pm}$0.07)	& 1.60 (${\pm}$0.02) \\
$f(M)$ (M$_{\odot}$)			& 0.0084 (${\pm}$0.0010)		& 0.0083 (${\pm}$0.0004)	\\
\hline
\end{tabular}
\end{center}
\end{table}

Utilizing the mass function formula \citep{torres10}:
\begin{equation}
\label{eqn2}
		f(M)  =  C(1 - e^2)^{3/2} K_1^3 P	,
\end{equation}
(where the constant $C  =  1.03615\times10^{-7}$ when $K_1$ is in km s$^{-1}$ and $P$ is in days) gives $f(M) = 0.0083$.  
This can be written as: 
\begin{equation}
\label{eqn3}
		\frac{q^3}{(1+q)^2} = \frac{f(M)}{M_1 \sin^3 i}		.  
\end{equation}
With the photometric value of $\sin i$ as 0.866 and a plausible estimate
for $M_1$ as 4 $M_{\odot}$, we derive $q = 0.16$, consistent with Table
\ref{LC_model}, as mentioned above.  This leads to the secondary being a
late K type dwarf with $V$-magnitude less than 1\% that of the primary
and thus not detectable spectroscopically.  The separation of the
components, using Kepler's 3rd law, turns out to be about 13.2
$R_{\odot}$, so the rotational velocity of the primary, if synchronized,
would be $\sim$124 km s$^{-1}$.  With the derived inclination, a
measured rotation speed of about 107 km s$^{-1}$ would then be expected.

\subsection{Rotational Velocities}
\label{sec3.2}

\begin{figure}
  \centering
	\includegraphics[scale=0.48]{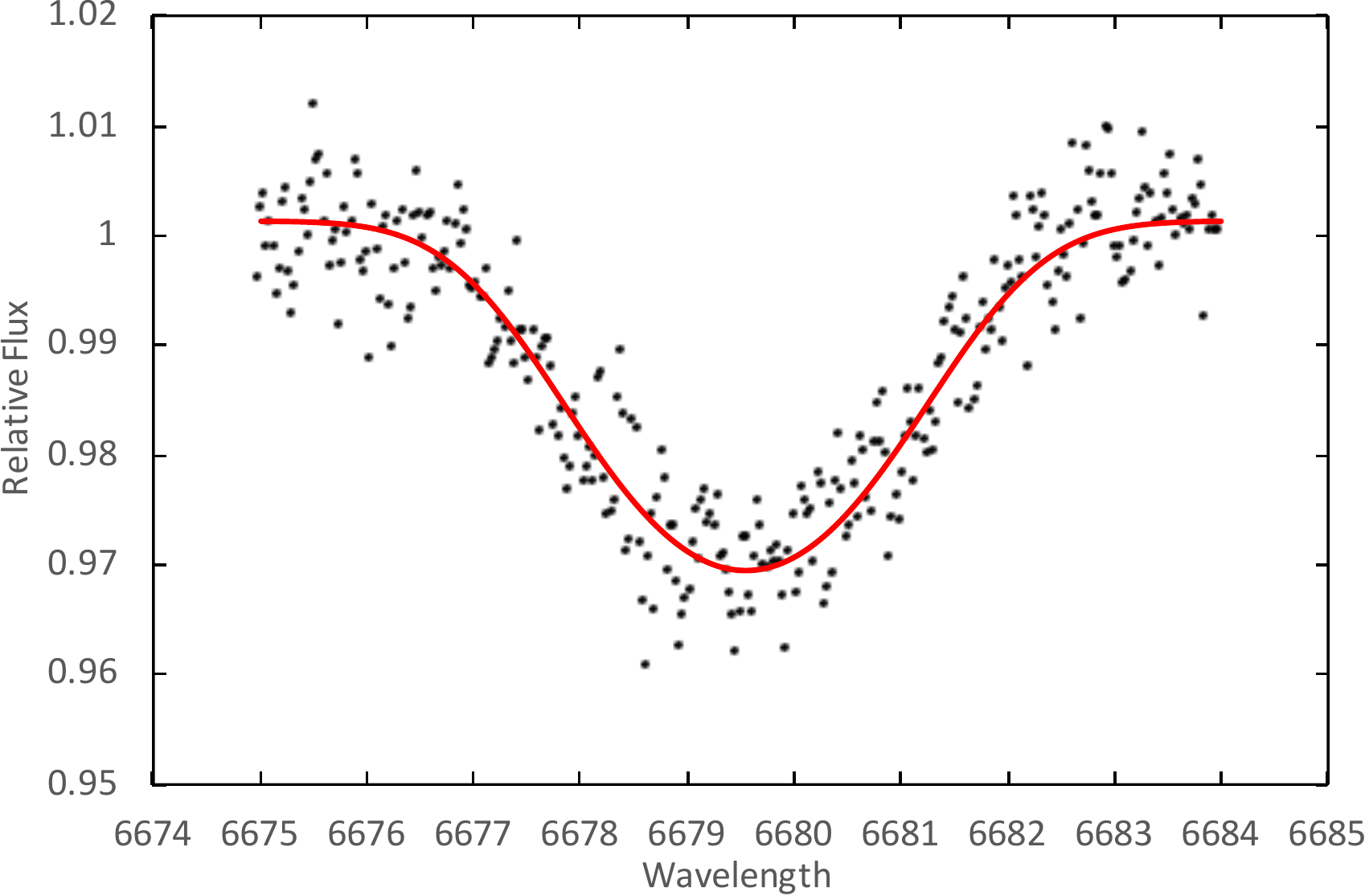}
	\includegraphics[scale=0.48]{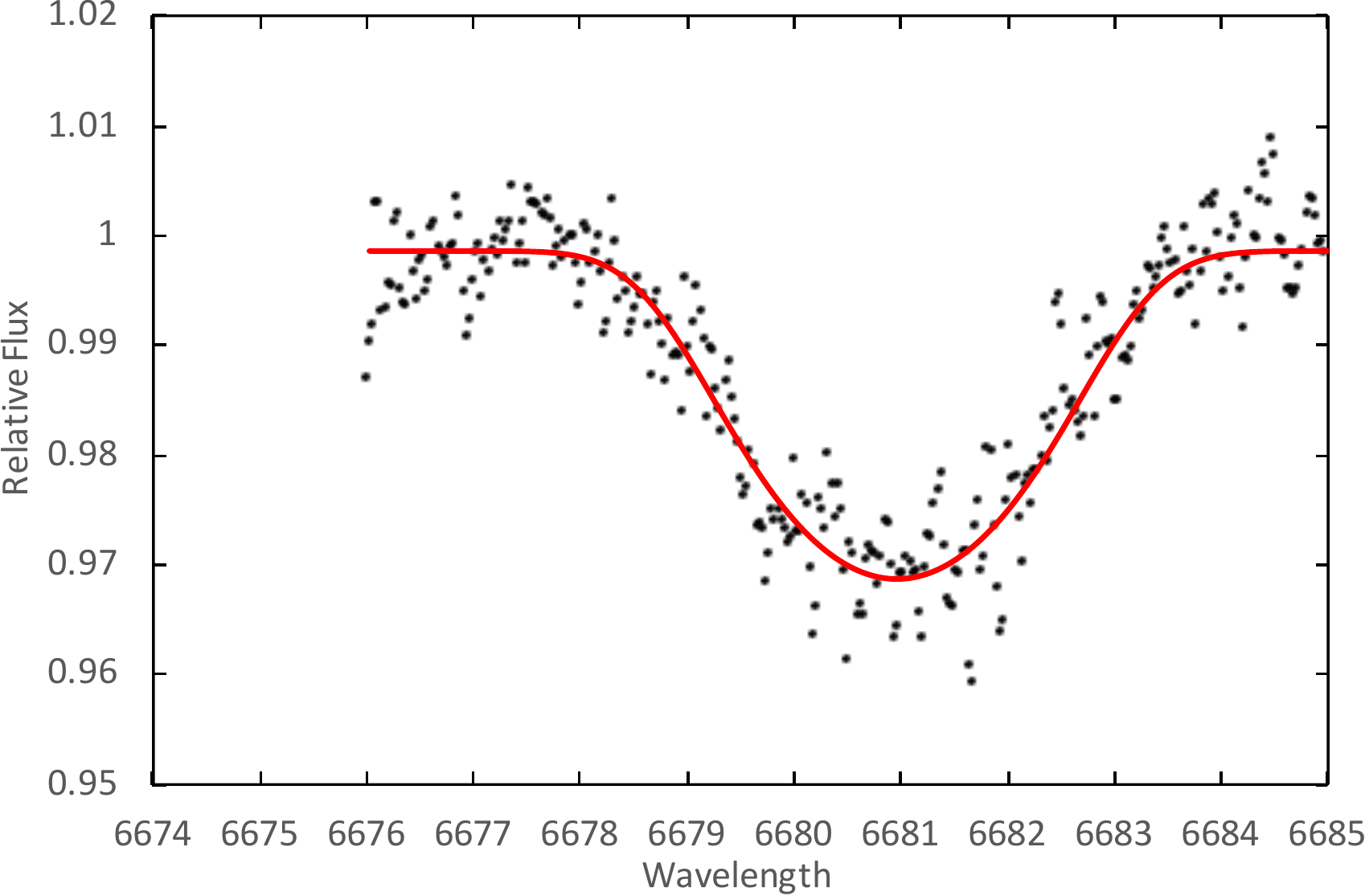}
\caption{Results of profile fitting to the He I 6678 lines at elongation phases 
(top panel for phase 0.21 and bottom panel for phase 0.73).
To show orbital motion and the line shift, the $x$-axis values are taken the same in both panels.
Wavelength is plotted on the horizontal axis, and relative flux on the vertical.}
\label{profit_fig}
\end{figure}

We fitted selected helium line profiles of PU Pup at various orbital
phases using the program {\sc prof} \citep{budding95,budding09}. {\sc
prof} convolves Gaussian and rotational broadening, and can characterize
the line profile in terms of the following adjustable parameters:
continuum intensity $I_c$, relative depth $I_d$ at mean wavelength
$\lambda_m$, rotational broadening parameter $r$ and Gaussian broadening
parameter $s$ for a given line, and a limb-darkening coefficient $u$.
Typical results of the profile fitting are shown in Figure
\ref{profit_fig} and given in Table \ref{profit}.

\begin{table}
\begin{center}
\caption{Profile Fitting Parameters For The He I 6678 Feature. 
\label{profit}
See section~\ref{sec3.2} for an explanation of the symbols. }
\begin{tabular}{lcc}
\hline
\multicolumn{1}{c}{Parameter}  & \multicolumn{1}{c}{Value} & 
\multicolumn{1}{c}{Error} \\
\hline
\multicolumn{3}{l}{Phase 0.21} \\
$I_c$ 		& 1.001			& 0.008	\\
$I_d$		& 0.027			& 0.007	\\
$\lambda_m$	& 6679.547		& 0.045 \\
$r$			& 2.31			& 0.20 \\
$s$			& 0.33			& 0.06 \\
$\chi^2/\nu$, $\Delta l$	& 1.23	& 0.01 \\
\hline
\multicolumn{3}{l}{Phase 0.73} \\
$I_c$ 		& 0.999			& 0.008	\\
$I_d$		& 0.025			& 0.010	\\
$\lambda_m$	& 6680.962		& 0.043 \\
$r$			& 2.27			& 0.22 \\
$s$			& 0.22			& 0.08 \\
$\chi^2/\nu$, $\Delta l$	& 1.26	& 0.01 \\
\hline
\end{tabular}
\end{center}
\end{table}

According to the value of $r$ (the rotational broadening parameter) in
Table \ref{profit}, the mean observed projected rotational velocity of
the primary component is $ 102 \: ({\pm} \: 20)$ km/s. Using absolute parameters
and assuming synchronous rotation, the theoretical projected rotational
velocity for the primary component was found to be $ 107 \: ({\pm} \: 10)$ km/s,
so synchronous rotation can be accepted within the assigned error
limits. The value of the Gaussian broadening $s$ changes between 25 and
15 km/s for the primary component according to the orbital phase. These
values could be related to  thermal broadening (with calculated thermal
velocities in the order of $\sim$8 km/s), together with turbulence
effects on the fairly distorted surface of the primary star.  
  
%------------------------------------------------------------------------------------------

\section{TESS Photometry}
\label{sec4}

%------------------------------------------------------------------------------------------
% Material on the HIPPARCOS fits has been removed at the referee's request

Light curve analysis of the TESS data has been carried out using 
{\sc WinFitter} program and the numerical
integration code of \citet{wilson71} combined with the Monte Carlo
({\sc wd+mc}) optimization procedure discussed by \citet{zola04}. This
second method allows modelling the binned TESS light curve of PU Pup by means  of
surface equipotentials, with suitable coefficients for gravity and
reflection effect parametrization. In this {\sc wd+mc} method, a range
of variation, fixed by physically feasible limits, is set for each
adjustable parameter. These ranges were selected in cognizance of the
{\sc WinFitter} modelling. The WD + MC
program follows a search method involving the solution space consisting of tens of thousands of
solutions in the given variation range of each adjustable parameter. Therefore, 
in the range of variation given in the manuscript, the WD + MC program performs mass
and inclination searches (q-search and i-search.)

{\sc WinFitter} has been applied in a number of recent similar 
studies to the present one. Its fitting function derives
from \citet{kopal59}'s treatment of close binary proximity effects, including tidal
distortions of stellar envelopes of finite mass, and luminous interaction factors
from the theoretical reflection-effect formulae of \citet{hosokawa58}. 
The main geometric parameters determined from the optimal
fitting of light-curve data with this model are the orbital inclination
$i$ and the mean radii ($r_{1, 2}$) given in terms of the mean
separation of the two components, corresponding to equivalent single stars.
The latest version of {\sc WinFitter} is available from \citet{rhodes20}.

The phases of the TESS photometric observations  were computed using our
derived new ephemeris, given as Equation \ref{eqn1} in Section
\ref{sec2}.  This was constructed from analysis of PU Pup's times of
minimum light, mentioned above. The phased TESS data were binned with
between 100 and 150 individual points per phase bin, allowing more
binned points at the eclipse phases. In this way, 128  points
together with their uncertainties were used in the analysis. The
uncertainties were derived from the published measurement information in
the source data.

Regarding the orbital inclination, we used the following basic formula:
\begin{equation}
\label{eqn_i}
		\delta ^{2} = cos^2 i + sin^2 i \ sin^2\phi,
\end{equation}
where $\delta$ is the separation of the two component star centres projected onto the celestial sphere, and $\phi$ is the orbital phase-angle and calculated from the primary 
mid-eclipse $T_{0}$ at time $T$ as $\phi = 2 \pi (T - T_{0})/P$. 
The term `grazing' means a very small, slight eclipse when the phase  $\phi$ $\sim 0$.
Then $cos \ i \approx \delta$.
Considering $\delta \approx r_{1} + r_{2}$ and taking $r_{1} + r_{2} \gtrsim 0.5$ from the preliminary curve-fits of \citet{budding19},
we find that the orbital inclination $i$ should be not far from 60$^{\circ}$.
 Thus, inclination limits were
set as $30^\circ < i <   70^\circ$. The input range of the zero phase
shift was taken as $-0.01 < \Delta \phi_0 <0.01$. The effective
temperature ($T_e$) of the primary component of PU Pup was estimated
from its spectral type (see Section \ref{sec1}). A compromise value of
11500 ($\pm \: 500$) K was adopted. In a similar way, a mass for the primary
was tentatively estimated as 4 M$_{\odot}$. Using the relationship
between mass function and mass ratio (Section \ref{sec3.1}, Equation
\ref{eqn3}), the feasible input range of mass ratio ($q = M_2/M_1$)
narrows to the range 0.15 --- 0.18. This implies a mid-K type MS dwarf for the
secondary with an effective temperature $T_e$ of around 5000 K. Since spectral features of the
secondary component are not visible (see Section \ref{sec3}) and the
photometric contribution of the secondary is small (Table \ref{LC_model}),
a direct confirmation of its properties is not practicable.

\begin{figure}
  \centering
	\includegraphics[scale=0.50]{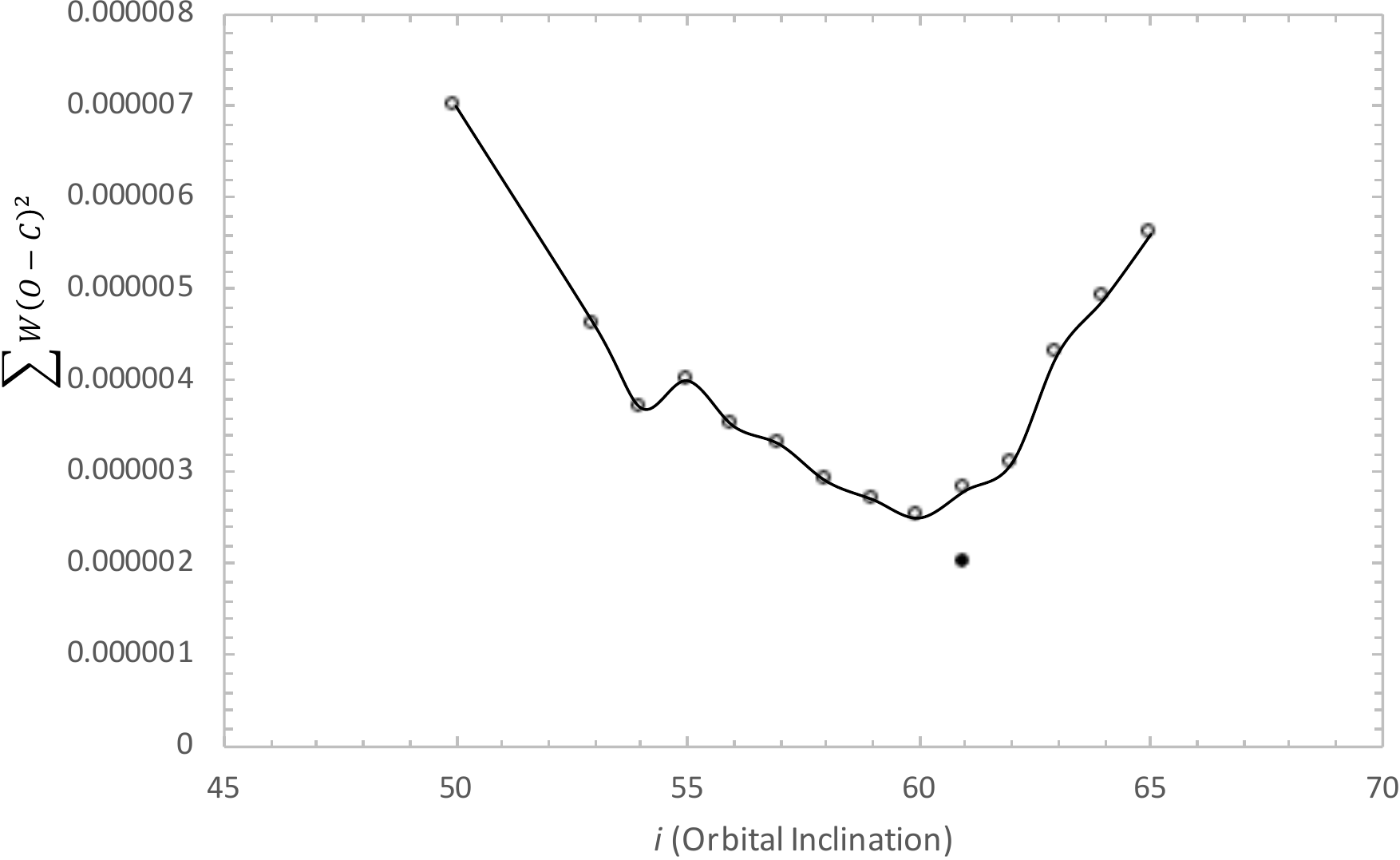}
	\includegraphics[scale=0.50]{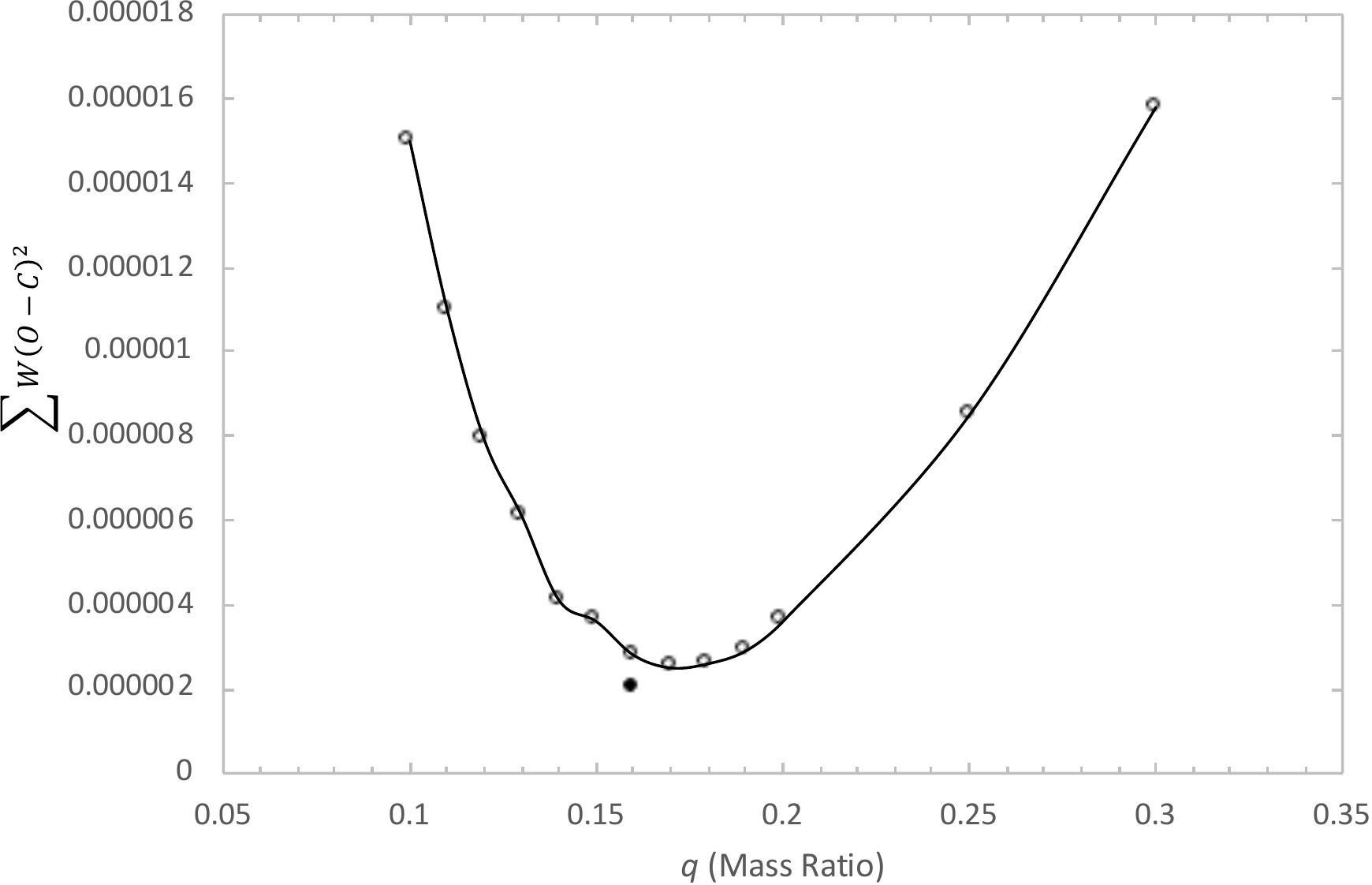}
\caption{Behavior of the weighted sum of the squared residuals, $\sum W(O - C)^2$, as a function of orbital inclination $i$ (top panel) and mass ratio $q$ (bottom panel). 
The values of $i = 61^\circ$ and $q = 0.16$ of the final model are also shown as filled circles.
}
\label{search}
\end{figure}

The available range for the non-dimensional surface potential parameters
$\Omega_1$ and $\Omega_2$  was set as the range 1.0 --- 8.0. The bolometric gravity
effect exponent ($\beta_{1}$) and  albedo ($A_{1}$) were taken as
empirical adjustables, noting that the effect of the ellipsoidal form of
primary component predominates. Given that the primary component has a
radiative envelope, the input range for $\beta_{1}$ was set as 0.5 ---
2. A range from 0.5 to 0.995 was allowed for the primary's fractional
luminosity ($L_{1}$). A quadratic limb-darkening law was used with
limb-darkening coefficients  taken from \citet{claret17}, where the
effective wavelength of the broad  TESS filter transmission function is
derived for given $T_e$ values. In the present case this is not far from
800 nm. It was assumed that the components of PU Pup are rotating
synchronously in a circular orbit.
On the other hand, an $i$-search was performed for the values of orbital 
inclination between 50 and 70 degrees and
a $q$-search for the values of the mass ratio between 0.10 and 0.30, using the DC code of the WD method. 
Fig. \ref{search} shows the weighted sum of squared residuals for ranges in the orbital inclination and mass ratio, $\sum W(O - C)^2$, with $O$ being the observed data and $C$ the calculated points.
As can be seen from these figure, the variation of
the weighted sum of the squared residuals versus orbital inclination and mass ratio
give a minimum around $i = 60^\circ$ and $q = 0.17$, respectively.
The values of $i = 61^\circ$ and $q = 0.16$ obtained in the final model match these minimum values.

\begin{figure}
  \centering
%  \vspace{6cm}
	\includegraphics[scale=0.48]{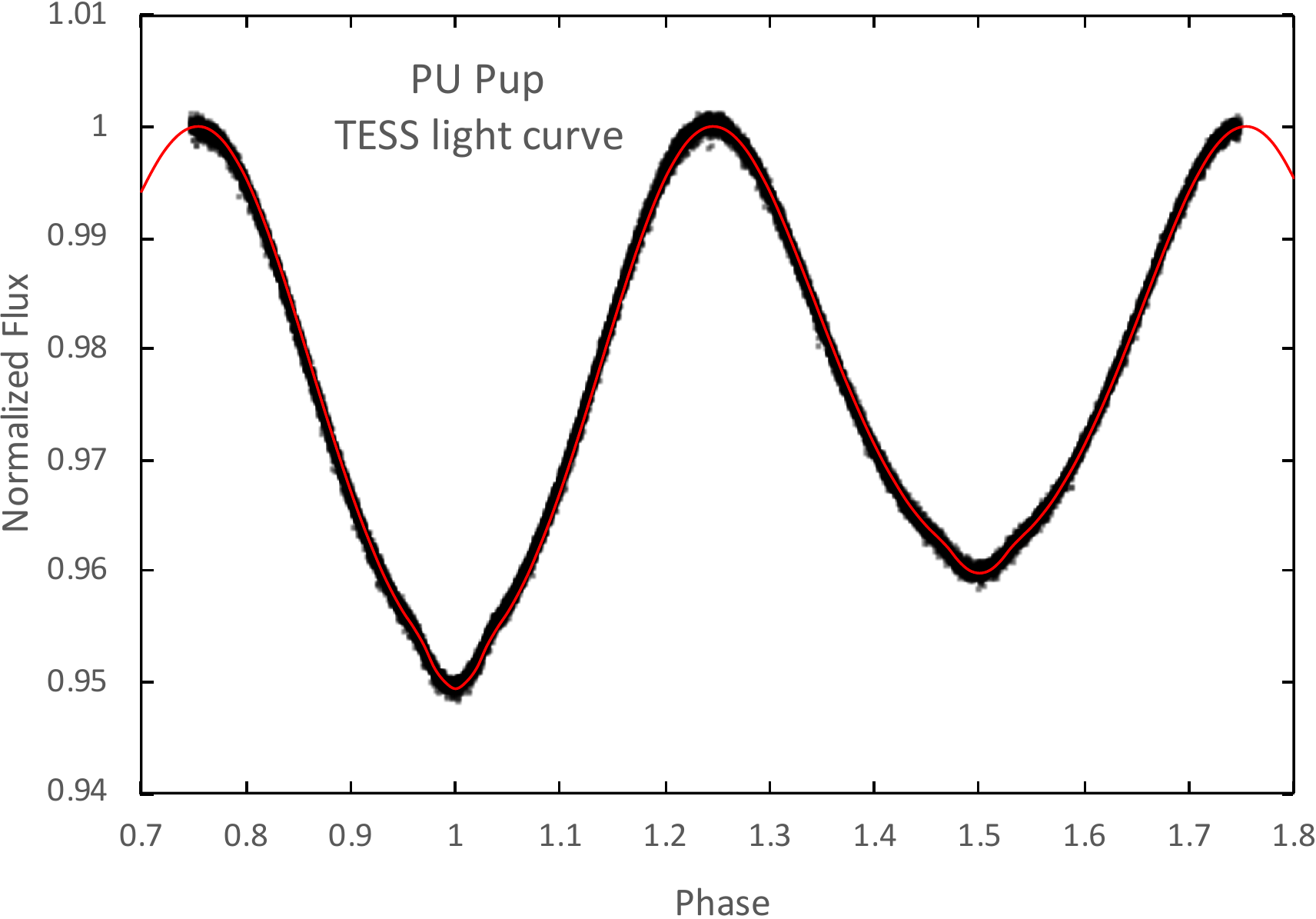}
\caption{TESS light curve of PU Pup (the points) and {\sc wd+mc} model (the red line) fitting.
Normalized flux is plotted on the vertical axis, where the total flux of the system is set to unity.  Orbital
phase is plotted on the horizontal axis, starting at 0.7 and going to 1.8.}
\label{tesslc}
\end{figure}

\begin{figure}
  \centering
%  \vspace{6cm}
	\includegraphics[scale=0.30]{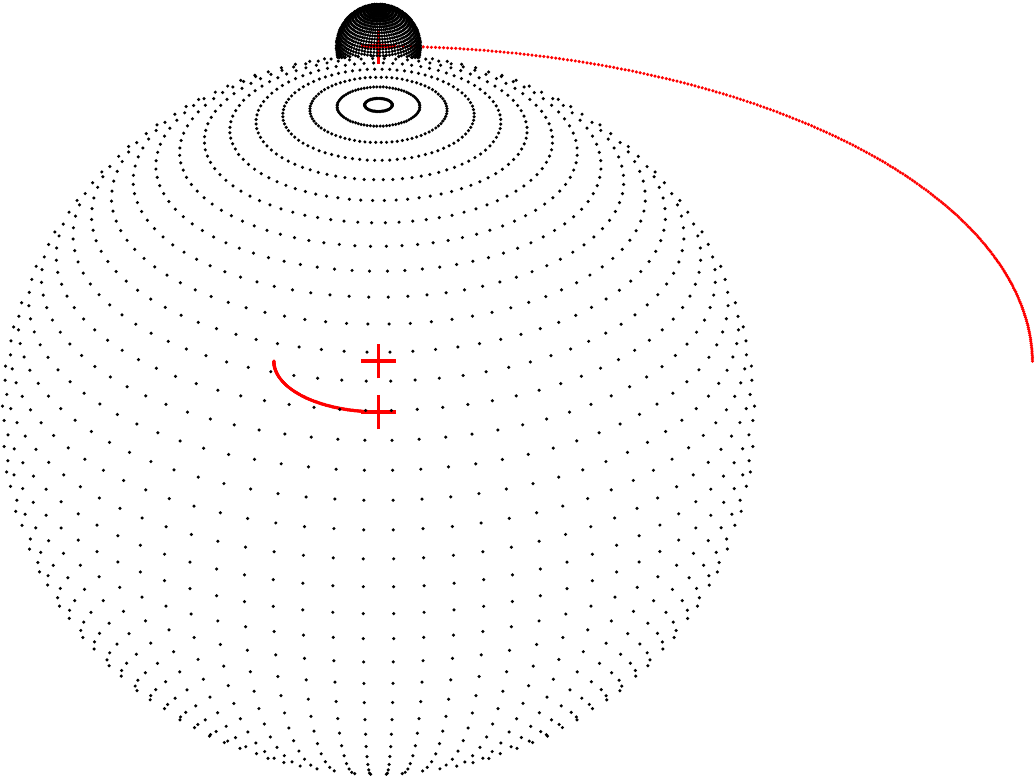}
	\includegraphics[scale=0.30]{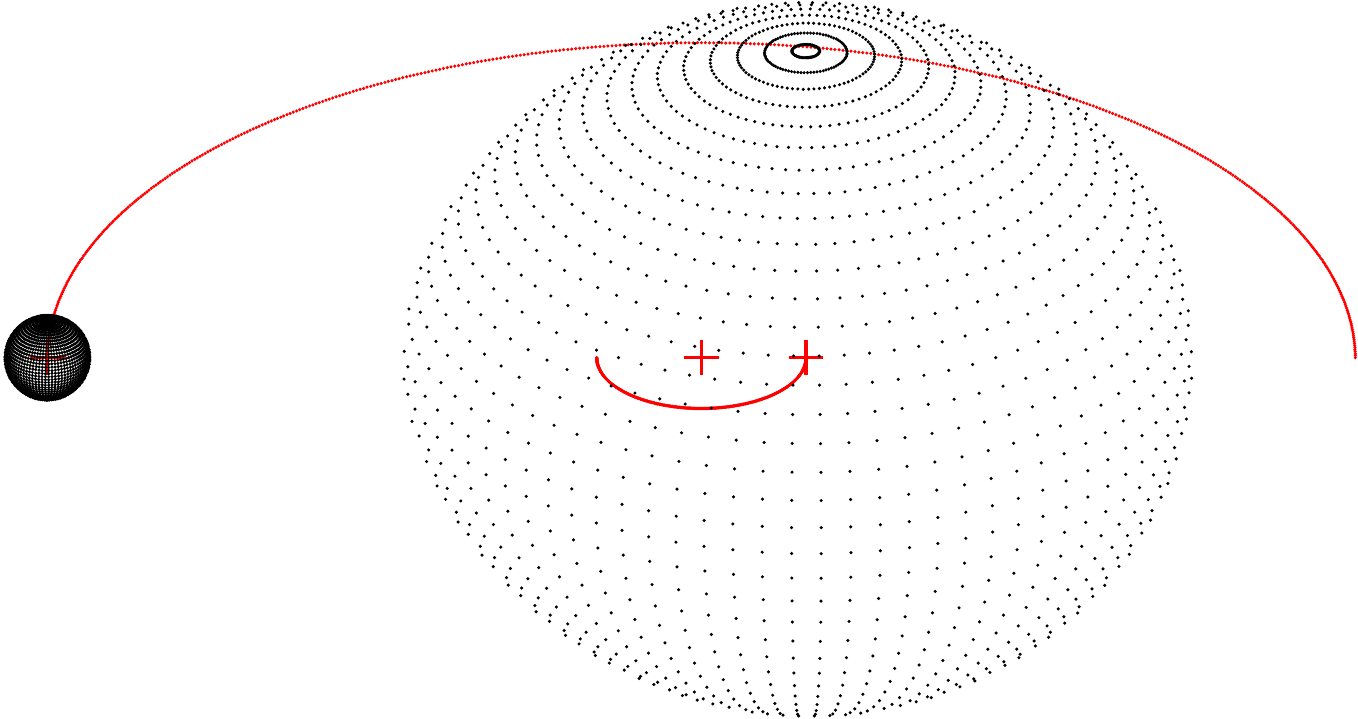}\\
	\includegraphics[scale=0.30]{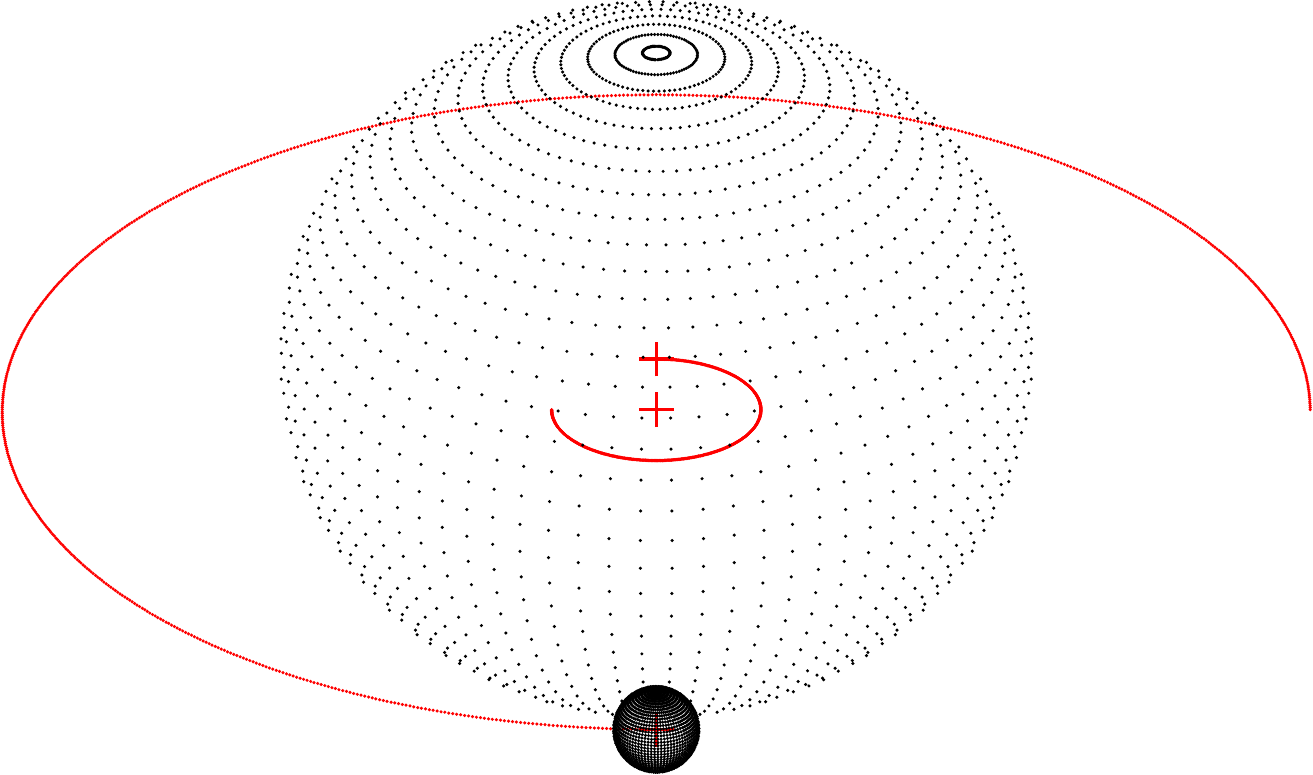}
	\includegraphics[scale=0.30]{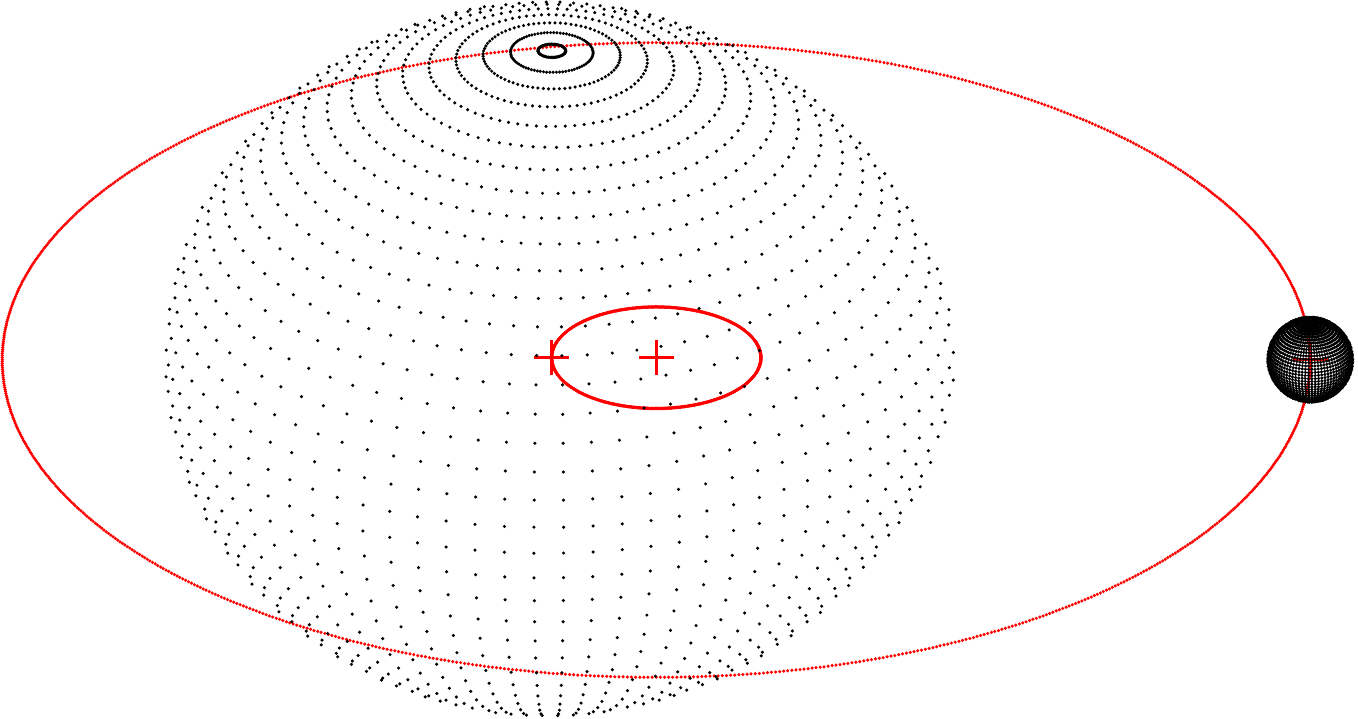}
\caption{3D model of PU Pup at conjunction  
(left bottom panel at phase 0.0 and left top panel at phase 0.5) and elongation 
(right bottom panel at phase 0.25 and right top panel at phase 0.75).}
\label{3D_model}
\end{figure}

The final parameters for the best-fitting {\sc wd+mc} model light curve
of PU Pup are given in Table \ref{LC_model}. The uncertainties of the
adjustables are given as output from the {\sc wd+mc} program and
correspond, formally, to a 90$\%$ confidence level \citep{zola04}.
$\chi^{2}$ was calculated as 
$\sum_i(\emph{l}_{i,o}-\emph{l}_{i,c})^{2}$/$\Delta\emph{l}_{i}^{2}$
from \citet{bevington69}, where  $l_{i,o}$ and $l_{i,c}$ are the
observed and calculated light levels at a given phase, respectively, and
$\Delta\emph{l}_{i}$  is an error estimate for the measured values of
$l_{i,o}$, taken to be 0.0002 in the relative flux. The reduced 
$\chi$-squared is given as $\chi^{2}_{red}$=$\chi^{2}/\nu$, where $\nu$
is the number of degrees of freedom of the data set \citep{bevington69}.
The adopted best-fit {\sc wd+mc} light curve for the TESS photometry is
shown in Figure \ref{tesslc}. A three-dimensional projected
illustration, including the grazing  eclipses, obtained with {\sc
BinaryMaker} \citep{bradstreet02}, is displayed in Figure
\ref{3D_model}.

The {\sc wd+mc} curve-fittings are compared with those of {\sc
WinFitter} ({\sc wf6}) in Table \ref{LC_model}. It can be seen directly
that there is a fair measure of agreement in the main geometric
parameters.  The fittings call for some degree of third light ($L_3$).
This related to how well the scale of the main ellipticity effect is
controlled by the assigned parameters.  This includes the proximity
effect coefficients that depend on the given wavelength and
temperatures.  In the present context it is the primary's tidal
distortion, related to $\beta_1$ ($\tau_1$), and perhaps the secondary's
`reflection effect', depending on $A_2$ ($E_2$), that can be
influential.

It is appropriate to say more about these coefficients and their effects
on the fittings. The $\beta$ parameters ({\sc wd+mc})represent the
bolometric index of the gravity darkening, i.e.\ $H/H_0 =
(g/g_0)^\beta$, where $H $ and $H_0$ are the local and mean surface
radiative fluxes; $g$ and $g_0$ are the corresponding local and mean
surface gravities. The $\tau$s are the corresponding  coefficients in
{\sc wf6}.  They represent the same indices, but corrected for the
effect of wavelength and temperature of observation: an operation
performed internally in both codes.  It can be seen from Table~4-5 in
Kopal's (1959) book that these corrections reduce the role of the
coefficient at higher temperatures and longer wavelengths.  But both
primary gravity coefficients $\beta_1$ and $\tau_1$ resulting from
optimal curve-fitting to the TESS photometry are in keeping with, or
somewhat greater than, the standard von Zeipel value of $\beta_1 = 1$
for a radiative envelope. The corresponding conversions for the
reflection coefficient $A_2$ or $E_2$ (sometimes called geometric
albedo) enhance the coefficient over its bolometric value at the low
temperature of the secondary, but the low relative size of the secondary
implies an essentially low scale of light reflection in PU Pup.

The difference in the zero phase correction $\Delta \phi_0$ arises from
the use of different ephemerides:  {\sc wf6} has used the period
provided by the HIPPARCOS Epoch Photometry Catalogue \citep{esa}
together with a preliminary time of minimum taken to correspond with the
lowest value of the TESS fluxes. {\sc wd+mc} used the new ephemerides
presented in the previous section.

The difference in the limb-darkening coefficients ($x_i$ and $y_i$ in
Table~6) arise mainly from a slight but significant difference in the
formulation between that of Claret (2017), used in {\sc wd+mc}, and
Kopal (1959 -- ch.\ 4), used in {\sc wf6}. The second-order
limb-darkening effects ($y$s) were fixed at low values in the {\sc wf6}
model, but it should be noted that in empirical curve-fitting there
would be a linear correlation between the first and second-order
coefficients that compromises their separate determinability.  The
selection of the limb-darkening approximation would have more
significance for PU Pup if the eclipses were more prominent.

\begin{table}
\begin{center}
\caption{The Results of Optimal Curve-fitting to the TESS Photometry of PU Pup 
using {\sc wd+mc} and {\sc wf6} Photometry Modelling Programs. 
There is reasonable agreement between these different fitting codes
that the inclination is close to 60$^{\circ}$;
the primary relative radius is $\sim$0.5 (and therefore the primary is not far from
contact with its Roche Lobe), and the primary star is about 6 times larger in radius
than its companion.  The secondary is involved in the very shallow eclipses ($\sim$0.5\%
 of the system light in depth), but other than that, its effects are weak.
 More details are given in the text.
\label{LC_model}} 
\begin{tabular}{lccc}
\hline
\multicolumn{1}{c}{Parameter}  & \multicolumn{1}{c}{{\sc wd+mc}} & \multicolumn{1}{c}{{\sc wf6}} & 
\multicolumn{1}{c}{Error estimate} \\
\hline
$T_1$ (K)			& 11500	& 11500 & 500 \\
$T_2$ (K) 			& 5560	& 5000	& 350 \\
$q=M_2/M_1$			& 0.16	& 0.16	& 0.01  \\
$L_1$ 				& 0.85	& 0.90	& 0.04 \\
$L_2$				& 0.02 	& 0.01 	& 0.005 \\
$L_3$				& 0.13 	& 0.10	& 0.03 \\
$\Omega_1$			& 2.303	& --- 	& 0.03 \\
$\Omega_2$			& 4.282	& ---	& 0.41 \\
$r_1$ (mean)		& 0.49	& 0.51 	& 0.02 \\
$r_2$ (mean)		& 0.06	& 0.07	& 0.01 \\
$i$ (deg)			& 61.0	& 56.2	& 1.1 \\
$\Delta \phi$ (deg)	& 0.00	& 2.25	& 0.06 \\	
$\beta_1$ / $\tau_1$	& 1.30	& 0.49	& 0.23 \\
$\beta_2$ / $\tau_2$	& 0.32 	& 0.92	& --- \\
$A_1$ /	$E_1$		& 0.51	& 0.17	& 0.12 \\
$A_2$ /	$E_2$		& 0.50 	& 1.9	& --- \\
$x_1$				& 0.124 & 0.29	& --- \\
$x_2$ 				& 0.378 & 0.60	& --- \\
$y_1$				& 0.260 & --0.02	& --- \\
$y_2$ 				& 0.213 & --0.03	& ---\\
$\Delta l$			& 0.0002	& 0.0002	& --- \\
$\chi^2/\nu$ 		&1.09 		&1.2	& --- \\
\hline
\end{tabular}
\end{center}
\end{table}

% Should we shift the phase down by one orbit, in the following chart?

\section{Absolute Parameters}
\label{sec5}

In the well-known `eclipse method' photometric and spectroscopic
findings are combined, making use of Kepler's third law,  to derive
absolute stellar parameters. If the photometric mass ratio and orbital
inclination in Table \ref{LC_model} are used in Eqn \ref{eqn3}, the mass
of the primary component is found to be $M_1$ = 4.10 ($\pm \: 0.20$)
M$_{\odot}$. The  mass of the secondary component, $M_2 \approx 0.65$
M$_{\odot}$, then follows. The average distance between components, $A$,
is calculated from Kepler's third law. The fractional radii of
components, $r_{1,2}$, obtained from the photometric curve-fits, lead to
the listed absolute radii, $R_{1,2}$. Surface gravities ($g$) are then
directly derived. Determination of the bolometric magnitude ($M_{bol}$)
and luminosity ($L$) of the component stars  requires the effective
temperatures that were taken from Table \ref{LC_model}. In the calculations, 
effective temperature \textit{T$_{e}$} = 5771.8 ($\pm$0.7) K, 
\textit{M$_{bol}$} = 4.7554 ($\pm$0.0004) mag, 
\textit{BC} = --0.107 ($\pm$0.020) mag, and \textit{g} = 27423.2 ($\pm$7.9) cm/s$^2$ were used 
for solar values \citep{pecaut13}. 

The absolute visual magnitude, M$_{V}$, involves the bolometric
correction formula, $BC$ = $M_{bol}$ -- $M_{V}$.   Bolometric
corrections for the components were taken from the tabulation of
\citet{flower96}, according to the assigned effective temperatures. The
photometric distance is calculated using the formula, $M_{V}$ = $m_{V}$
+ 5 -- 5log$(d)$ -- $A_{V}$. For this, the interstellar absorption and
intrinsic color index were  computed using the method given by
\citet{tuncel16}. Firstly, the total absorption towards PU Pup in the
galactic disk in the $V$ band, $A_{\infty}(V)$, was taken from
\citet{schlafy11}, using the NASA Extragalactic Database\footnote{
http://ned.ipac.caltech.edu/forms/calculator.html}. Then the
interstellar absorption for PU Pup's distance, $A_{d}(V)$, was derived
from the formula given by \citet{bahcall80} (their Eqn.~8), using the
Gaia DR2 parallax \citep{gaia18}. The color excess for the system at
this distance, $d$, was estimated as $E_{d}(B - V)$ = $A_{d}(V)$/3.1.
Thus, the intrinsic color index of PU Pup was calculated as ($B -
V$)$_0$ = --0.13 mag. This photometric distance -- correcting for
interstellar absorption -- is then found to be 186 ($\pm$20) pc.

Our absolute  parameters for the PU Pup system are given in Table
\ref{abs_par} with their standard errors. By comparison, previous
photometric parallax calculations \citep{popper98,budding07} result in a
distance of 190 ($\pm$22) pc.  The astrometric parallaxes of  Gaia DR2
\citep{gaia18} and HIPPARCOS  \citep{vanleeuwen07} produce distances of
176 ($\pm$10) and 190 ($\pm$10) pc, respectively. This consistency between
the distances calculated by different methods, taking into account their
standard errors, allows confidence in the observationally determined
absolute parameters of PU Pup in the present study.

\begin{table}
\caption{Absolute Parameters of PU Pup.
Certain priors in this table were adopted from
published sources thus: $^{a}$ \citet{fabricius02}, $^{b}$ \citet{gaia18}, and $^{c}$  \citet{vanleeuwen07}. Terms are as defined in the text.
} \label{abs_par}
\small
\begin{center}
\begin{tabular}{lcc}
\hline
Parameter		& Primary		& Secondary	\\
\hline 
$A$ (R$_{\odot}$)	& \multicolumn{2}{c}{13.30 ($\pm$0.40)}		\\
$M$ (M$_{\odot}$)	& 4.10 ($\pm$0.20)	 & 0.65 ($\pm$0.05)	\\
$R$ (R$_{\odot}$)	& 6.60( $\pm$0.30) & 0.90 ($\pm$0.10)	\\
log $g$ ($cm/s^{2}$) & 3.40 ($\pm$0.05) & 4.30 ($\pm$0.10)	\\
$T$ (K)			& 11500 ($\pm$500) & 5000 ($\pm$350)		\\
$L$ (L$_{\odot}$)	& 695 ($\pm$80) & 0.50 ($\pm$0.10)	\\
$M_{bol}$ (mag)	& --2.35 ($\pm$0.20) & 5.54 ($\pm$0.40)	\\
$M_{V}$ (mag)		& --1.77 ($\pm$0.22) & 5.85 ($\pm$0.45)	\\
$E$($B - V$) (mag) &  \multicolumn{2}{c}{0.041}				\\
$B - V$ (mag)		& \multicolumn{2}{c}{--0.09$^{a}$}	\\
$V$ (mag)		& \multicolumn{2}{c}{4.67$^{a}$}	\\
$M_{V}$ $(system)$ (mag)	& \multicolumn{2}{c}{--1.80 ($\pm$0.25)}	\\
$d$ (pc)			& \multicolumn{2}{c}{186 ($\pm$20)}				\\
$d_{Gaia-DR2}$(pc)		& \multicolumn{2}{c}{176 ($\pm$10)$^{b}$}		\\
$d_{HIP}$(pc)		& \multicolumn{2}{c}{190 ($\pm$10)$^{c}$}			\\
\hline
\end{tabular}
\end{center}
\end{table}

\section{Evolutionary Status}
\label{sec6}

The ellipticity effect predominantly from the primary component and 
small (grazing) eclipses give the ratio of radii of components as 0.15
in the light curve solutions. 
In other words, the radius of primary
component is approximately 7 times larger than that of secondary
component, and, as indicated by the low value of $\Omega_1$ in Table
\ref{LC_model}, this star is relatively close to filling its Roche Lobe. 
 That is, the primary and secondary components fill 93\% and 50\% of their
Roche lobes. This unusual configuration raises a challenge in piecing together the
age and evolution of the system.

The Padova evolution models \citep{mar} display theoretical
(mass -- radius) and (mass -- $T_{e}$) isochrone locations in Figure
\ref{evol_fig1} corresponding to a metallicity of Z = 0.014. The primary
appears to be close to the Terminal Age of the Main Sequence for its
mass, with an age of about 170 My for Z = 0.014. However, it may be
noted that this evolution model predicts a smaller radius and lower
surface temperature than is observed for the secondary star.

A sensitive way to estimate Z from theoretical stellar models is from
the log ($T_{e}$) -- log ($g$) diagram. Looking at this in Figure
\ref{evol_fig3}, we deduce  that the metallicity and age of the primary
star are about Z = 0.014 ($\pm0.003$) and 170 ($\pm20$) {Myr. We show only the
primary in this diagram since the secondary is discordant for reasons
that we consider below. Other predictions (e.g. from the BaSTI and
Geneva stellar evolution models) were investigated and similar results
were found.

With regard to the apparent luminosity excess of the secondary, this
kind of problem is not uncommon in short-period binary systems
\citep[e.g.,][]{garrido19}. Such inflated radii for relatively low mass
components have been discussed in terms of fast rotation and tidal
effects  \citep[e.g.,][]{chabrier07,kraus11}. We should also keep in
mind that the luminosity of the primary is greater than that of the
secondary by about three orders of magnitude.

Padova models for a star of 0.65 M$_{\odot}$ with a typical young star
metallicity (Z = 0.014-19) give a radius of $\sim$0.6 R$_{\odot}$ at the
estimated age of around 170 My. These models show that even at $\sim$170
My, the low mass star is still condensing towards the Zero Age Main
Sequence, but the increased luminosity from this would account for less
than 1\% of an increase in radius, i.e.\ insufficient to explain the
$\sim$30\% increase in radius over a standard model for an unevolved
star of the secondary's mass. A rough estimate shows the amount of
radiated energy intercepted by the dwarf from the subgiant to be
$\sim$10 times its own inherent luminosity. Assuming that the mean
effective temperature of the secondary would increase by $\sim$20\% with
this energy input, the radius should double, in order to radiate away
the excess energy and remain in thermal equilibrium. Of course, some of
the flux received at the secondary may be simply scattered or else drive
kinetic mass motions in the dwarf's envelope, but in any case a
significant increase in radius of the secondary can be reasonably
anticipated on the basis of the heat received from the primary.

\begin{figure}
  \centering
	\includegraphics[scale=0.55]{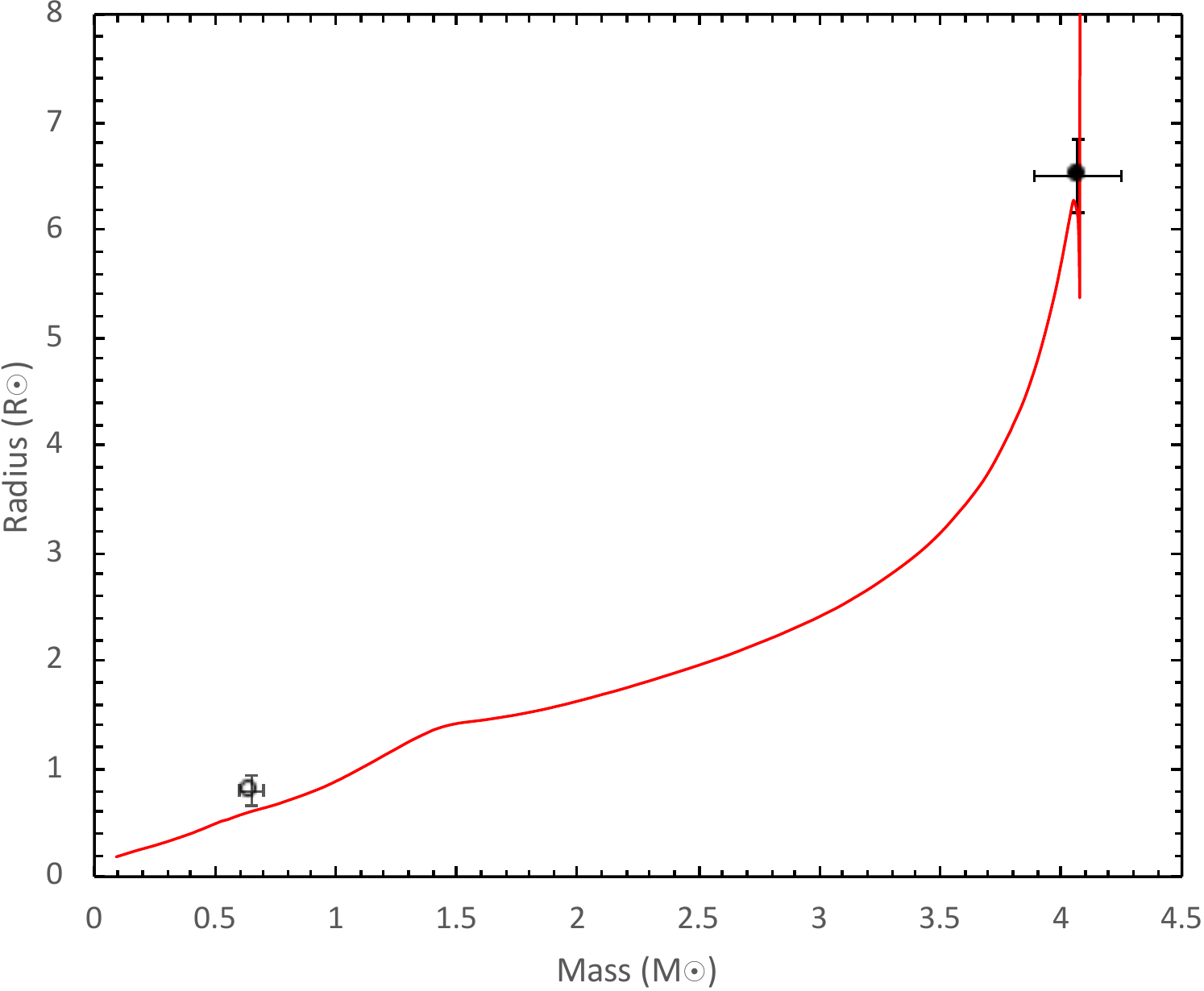}
	\includegraphics[scale=0.55]{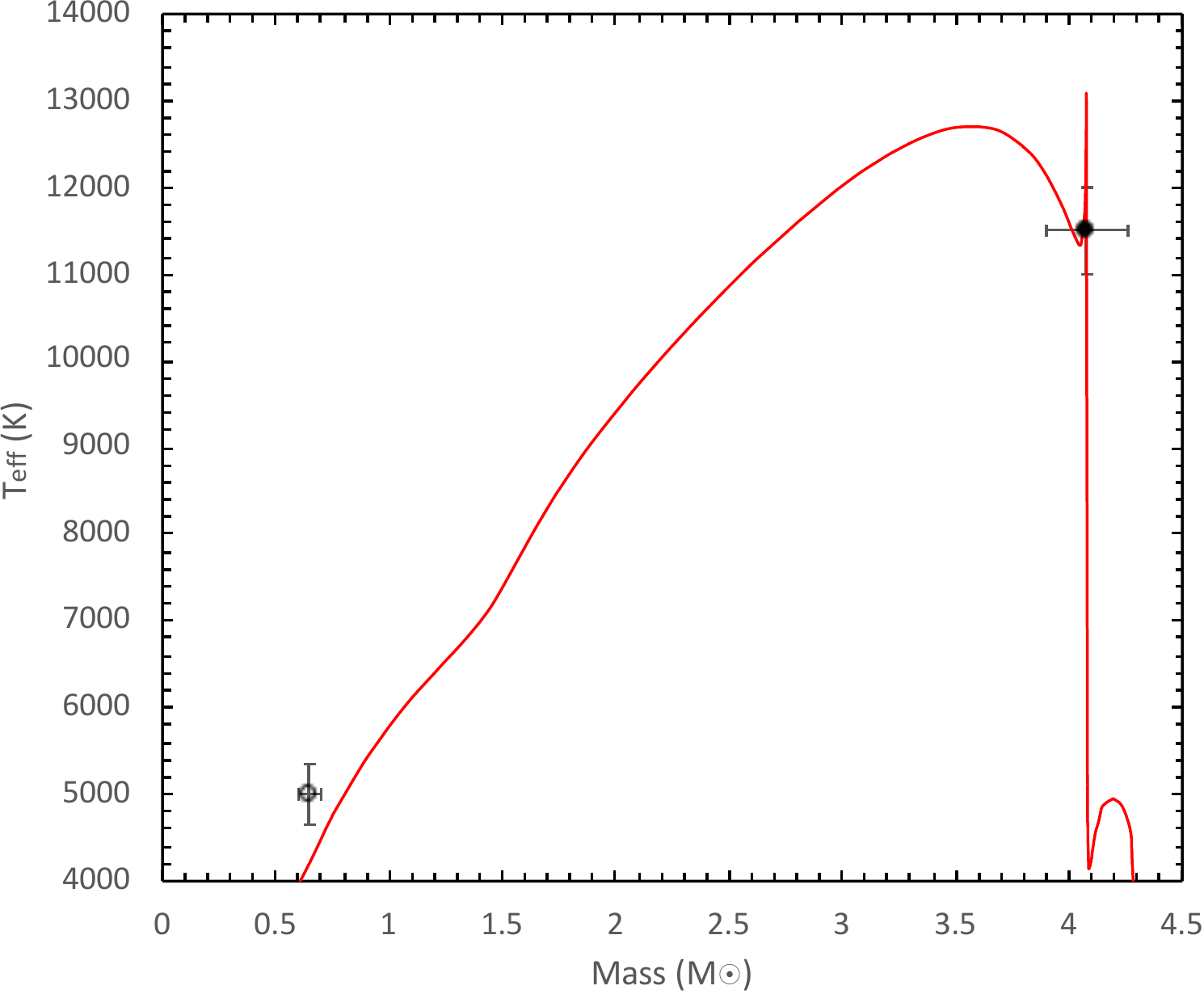}
\caption{Locations of the components of PU Pup in the mass-radius diagram (top panel) 
and mass--$T_{\rm eff}$ (effective temperature) diagram (bottom panel). 
The Padova isochrone line of 170 My for Z = 0.014 \citep{bressan12} 
is indicated by the red line.  The filled and open circle symbols represent
primary and secondary components, respectively. 
Vertical and horizontal lines show error bars of the measured quantities.
}
\label{evol_fig1}
\end{figure}

\begin{figure}
  \centering
	\includegraphics[scale=0.55]{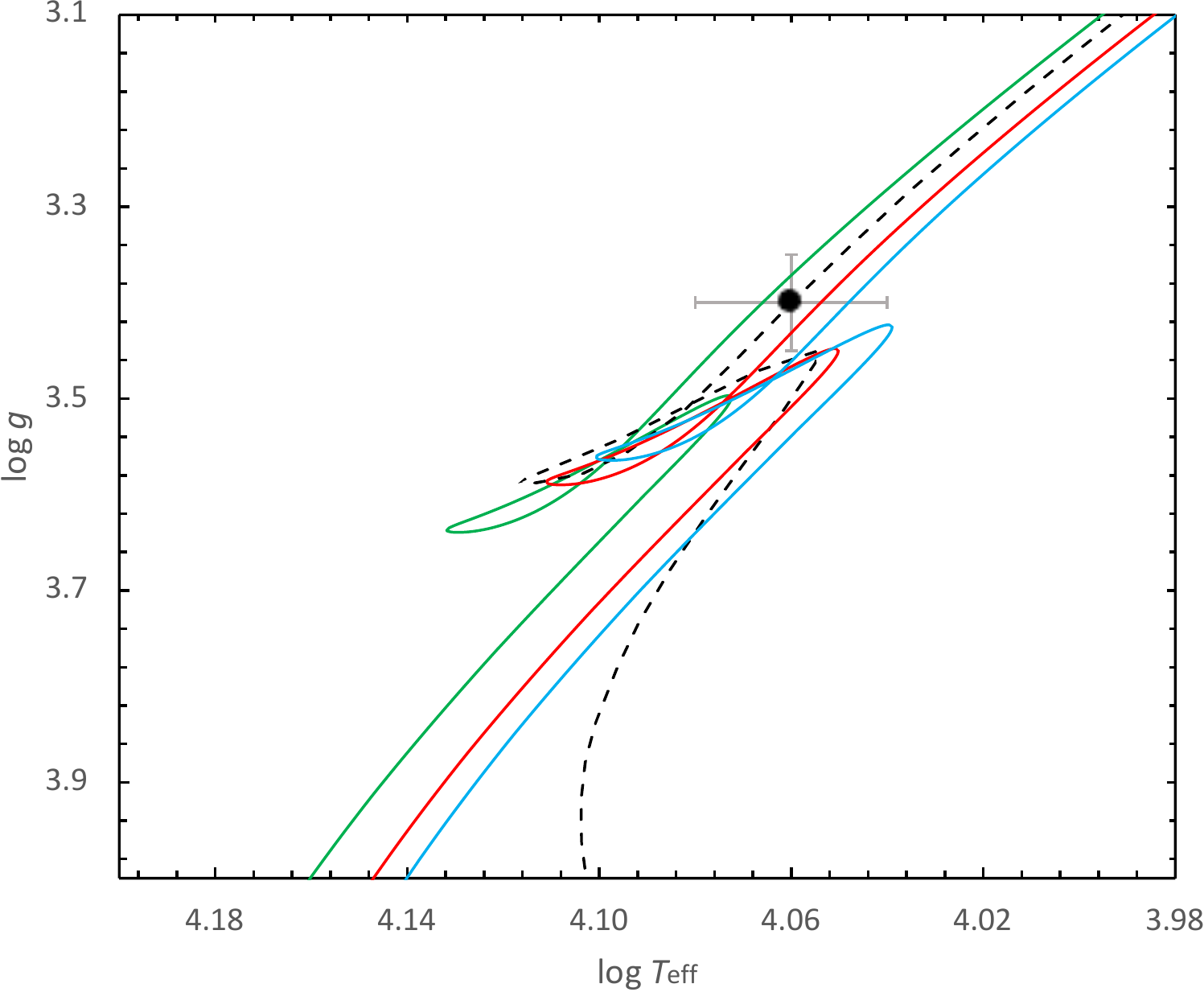}
\caption{Location of the primary component of PU Pup in the $\log T_{\rm eff} - \log g$ diagram. 
The Padova evolutionary tracks \citep{bressan12} for its mass are plotted for Z = 0.010  
(green line), 0.014 (red line) and 0.017 (blue line). 
The best-fitting isochrone of 170 Myr for Z = 0.014 is also indicated by the black dotted line.}
\label{evol_fig3}
\end{figure}

\section{Concluding Remarks}
\label{sec7}
 
PU Pup is an extraordinary close binary system. From historic data,
including that from the HIPPARCOS satellite, the light curve, apparently
of EB ($\beta$ Lyrae) kind, suggested either a close early-type pair
with a strong third light contribution, or an ellipsoidal variable. 
A preliminary fitting to the HIPPARCOS data  yielded a Main Sequence pair
containing a massive star with a third light of almost $50\%$ in the
system (thus resembling the system V454 Car). 
However, our spectral observations of PU Pup in 2008, 2014 and 2015 ruled out such a high
level of third light. Neither the spectrum of such a source nor that of
the secondary is visible.

%\add command failed again --- remember to remove the color statement in final draft

Further analysis of the HIPPARCOS data, together with the  RV
information, presented the secondary as a late-type Main Sequence dwarf,
consistent with a low mass ratio {\color{blue} \citep[for details, see][]{budding19}).}
This model pointed to a low
inclination, with the elliptical form of the primary component producing
the dominant effect rather than eclipses.  The light contribution of the
secondary was estimated at less than a few percent of the total light. 

The TESS satellite, launched in 2018, included 9 continuous cycles of PU
Pup's photometric light curve in its Sector~7 data. These very high
accuracy observations reveal interesting small eclipse effects at the
bottom of the light minima. The discovery of these small eclipses allows
significant progress in uncovering the system's parameters. The grazing
eclipses constrain the system's orbital inclination to be around 60
degrees. According to our revised model for the light and RV curves, PU
Pup is approaching a semi-detached binary configuration, where it is the
more massive primary that is close to filling its Roche limiting surface
\citep[cf.][Chapter~7]{kopal59}. This binary is thus moving towards the
rarely seen `fast phase' of interactive evolution. 

Given the primary almost filling its lobe together and its over-sized secondary, 
PU Pup may be considered a near-Algol system.  It could be considered similar to S Equulei (\citealt{plavec}, \citealt{Qian1}), BH Virginis (\citealt{tian}, 
\citealt{zeilik}, \citealt{zhu}), 
EG Cephei (\citealt{Erdem}, \citealt{zhu1}), and GQ Draconis (\citealt{atay}, \citealt{Qian2}). 
These systems have been reported to show signs of period variation.  In some 
cases, these may be close to the end of a mass-transferring (RLOF) phase, 
and this might be the case for PU Pup.  We are grateful to the unnamed referee for pointing out these comparable cases. This unusual system should therefore continue
to be monitored closely for evidence of period variation or other indications of instability.

\section{Acknowledgements}
 
Generous allocations of time on the 1m McLennan Telescope and HERCULES
spectrograph at the Mt John University Observatory in support of the
Southern Binaries Programme have been made available through its TAC and
supported by its  Director, Dr.\ K.\ Pollard and previous Director,
Prof.\ J.\ B.\ Hearnshaw. Useful help at the telescope were provided by
the MJUO management (N.\ Frost and previously A.\ Gilmore~\& P.\
Kilmartin). Considerable assistance with the use and development of the
{\sc hrsp} software was given by its author Dr.\ J.\ Skuljan, and very
helpful work with initial data reduction was carried out by R.\ J.\
Butland. We thank the anonymous referee for their guidance, which
led to an improved paper.

General support for this programme has been shown by the the School of
Chemical and Physical Sciences of the Victoria University of Wellington,
as well as the \c{C}anakkale Onsekiz Mart University, Turkey, notably
Prof.\ O.\ Demircan. We thank the Royal Astronomical Society of New Zealand,
particularly its Variable Stars South section
(http://www.variablestarssouth.org), for support.

It is a pleasure to express our appreciation of the high-quality and
ready availability, via the Mikulski Archive for Space Telescopes
(MAST), of data collected by the TESS mission. Funding for the TESS
mission is provided by the NASA Explorer Program. This research has made
use of the SIMBAD data base, operated at CDS, Strasbourg, France, and of
NASA's Astrophysics Data System Bibliographic Services. We thank the
University of Queensland for their assistance of collaboration tools.

\section{Data availability}
All data included in this article are available as listed in the paper or
from the online supplementary material it cites.

{}

%%%%%%%%%%%%%%%%% APPENDICES %%%%%%%%%%%%%%%%%%%%%

\appendix
\label{appendix}

\pagebreak
\section{Table of Minima Times of PU Pup}
\label{appendix_a}

%-------------------------------------------------tableA1
\setcounter{table}{0}
\renewcommand{\thetable}{A\arabic{table}}
\begin{table*}[!htbp]
	\centering
	\caption{Times of Minima of PU Pup.}
	\label{tableA1}
	\begin{tabular}{ccccc} % four columns, alignment for each
		\hline
Time of Minimum	& Uncertainty	& Minimum	& Reference	& Remarks \\
(BJD)			&			& Type		&			&		  \\
\hline
2448060.0709	&	0.0017	&	Min I	&	HIPPARCOS	&	(a)	\\
2448061.3621	&	0.0020	&	Min II	&	HIPPARCOS	&	(a)	\\
2448501.6288	&	0.0022	&	Min I	&	HIPPARCOS	&	(a)	\\
2448502.9201	&	0.0020	&	Min II	&	HIPPARCOS	&	(a)	\\
2457033.3148	&	0.0030	&	Min I	&	KWS I	&	(a)	\\
2457034.6100	&	0.0020	&	Min II	&	KWS I	&	(a)	\\
2457800.2415	&	0.0020	&	Min I	&	KWS V	&	(a)	\\
2458134.6365	&	0.0025	&	Min II	&	KWS V	&	(a)	\\
2458492.2724	&	0.0002	&	Min I	&	TESS	&	(b)	\\
2458493.5724	&	0.0003	&	Min II	&	TESS	&	(b)	\\
2458494.8609	&	0.0002	&	Min I	&	TESS	&	(b)	\\
2458496.1505	&	0.0002	&	Min II	&	TESS	&	(b)	\\
2458497.4410	&	0.0001	&	Min I	&	TESS	&	(b)	\\
2458498.7320	&	0.0001	&	Min II	&	TESS	&	(b)	\\
2458500.0263	&	0.0001	&	Min I	&	TESS	&	(b)	\\
2458501.3156	&	0.0002	&	Min II	&	TESS	&	(b)	\\
2458502.6036	&	0.0002	&	Min I	&	TESS	&	(b)	\\
2458505.1885	&	0.0002	&	Min I	&	TESS	&	(b)	\\
2458506.4824	&	0.0002	&	Min II	&	TESS	&	(b)	\\
2458507.7658	&	0.0001	&	Min I	&	TESS	&	(b)	\\
2458509.0614	&	0.0002	&	Min II	&	TESS	&	(b)	\\
2458510.3554	&	0.0002	&	Min I	&	TESS	&	(b)	\\
2458511.6458	&	0.0002	&	Min II	&	TESS	&	(b)	\\
2458512.9341	&	0.0002	&	Min I	&	TESS	&	(b)	\\
2458514.2267	&	0.0004	&	Min II	&	TESS	&	(b)	\\
2458515.5190	&	0.0001	&	Min I	&	TESS	&	(b)	\\
\hline
\end{tabular}
 
\noindent (a) These times of minima were derived using the theoretical LC templates.

\noindent (b) These times of minima were obtained from photometric observations directly.

\end{table*}

%------------------------------------------------------------------------------------------------------------

\newpage
\section{Example Spectrum of PU Pup}
\label{appendix_b}
%-------------------------------------------------figureB1
\renewcommand{\thefigure}{B\arabic{figure}}
 \begin{center}
	\includegraphics[scale=0.41]{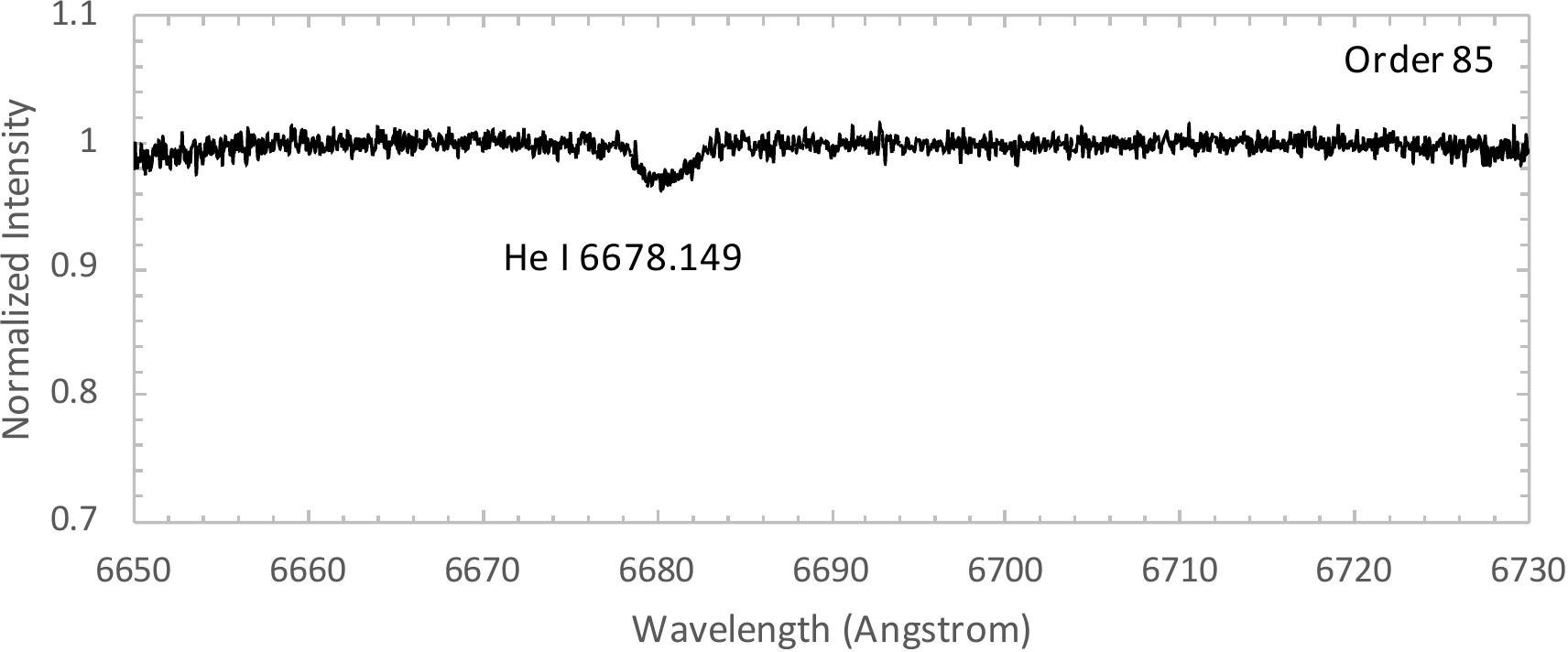}
	\includegraphics[scale=0.41]{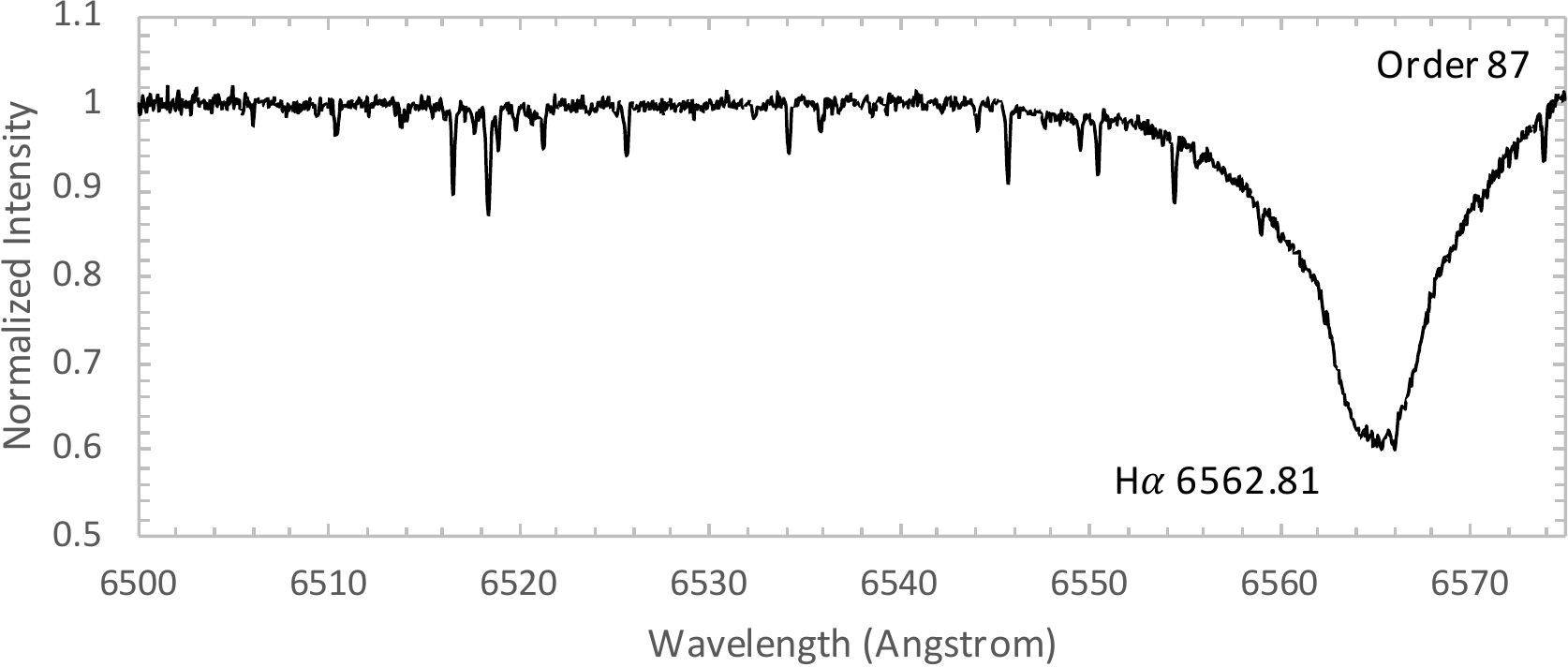} \\
		\includegraphics[scale=0.41]{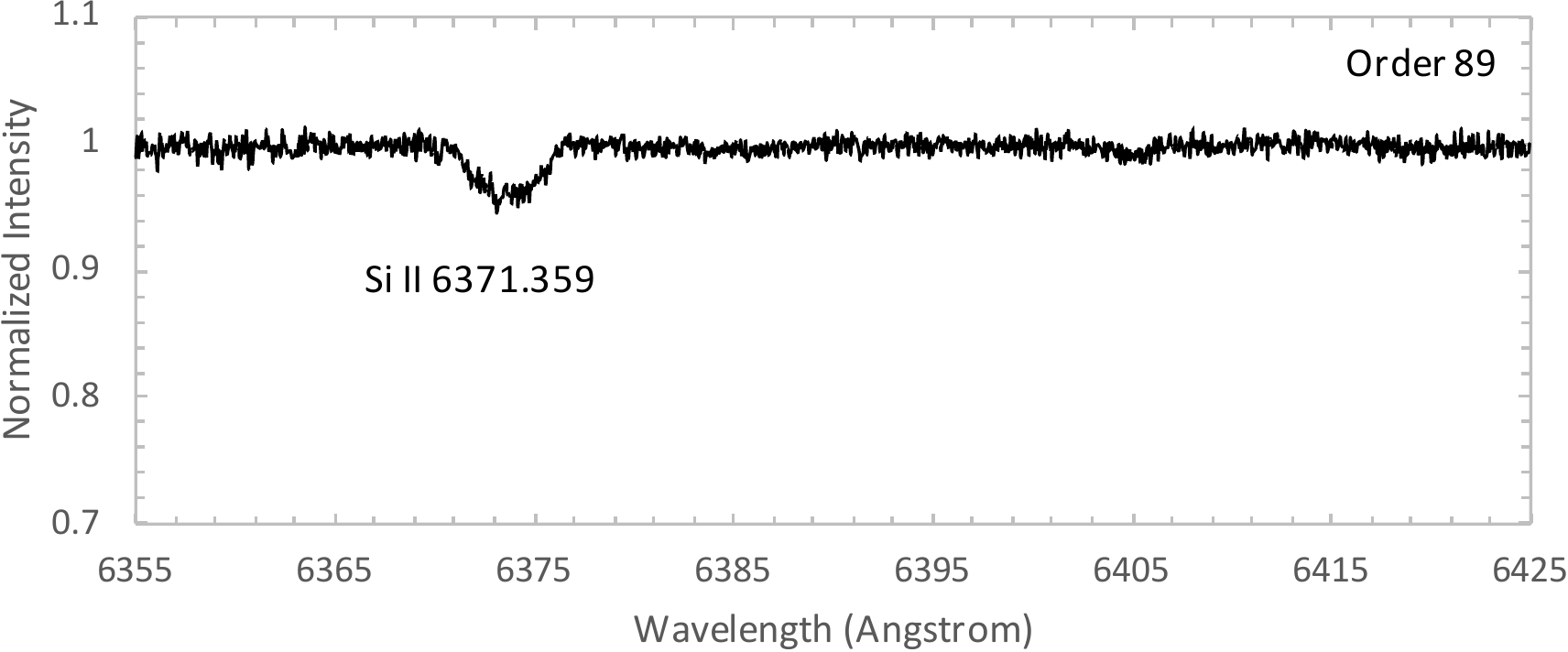}
		\includegraphics[scale=0.41]{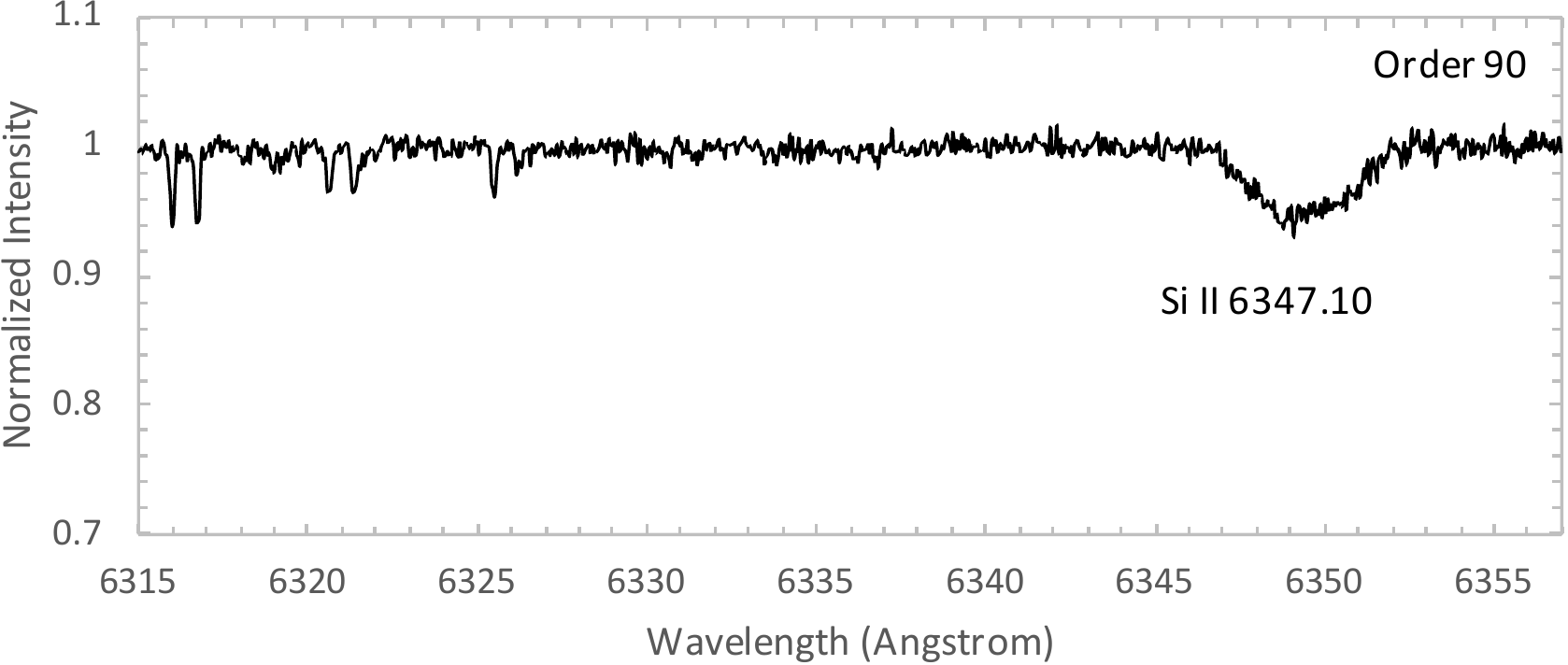} \\
	\includegraphics[scale=0.41]{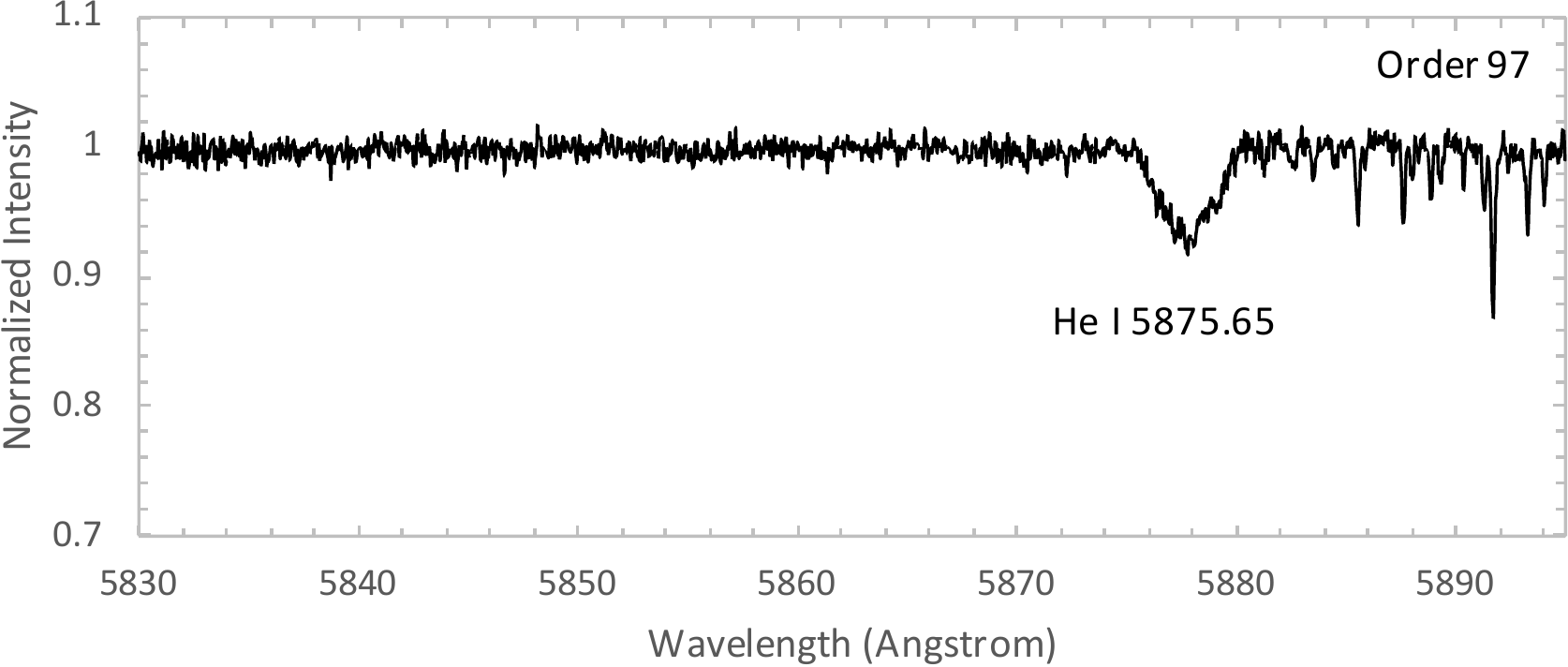}
	\includegraphics[scale=0.41]{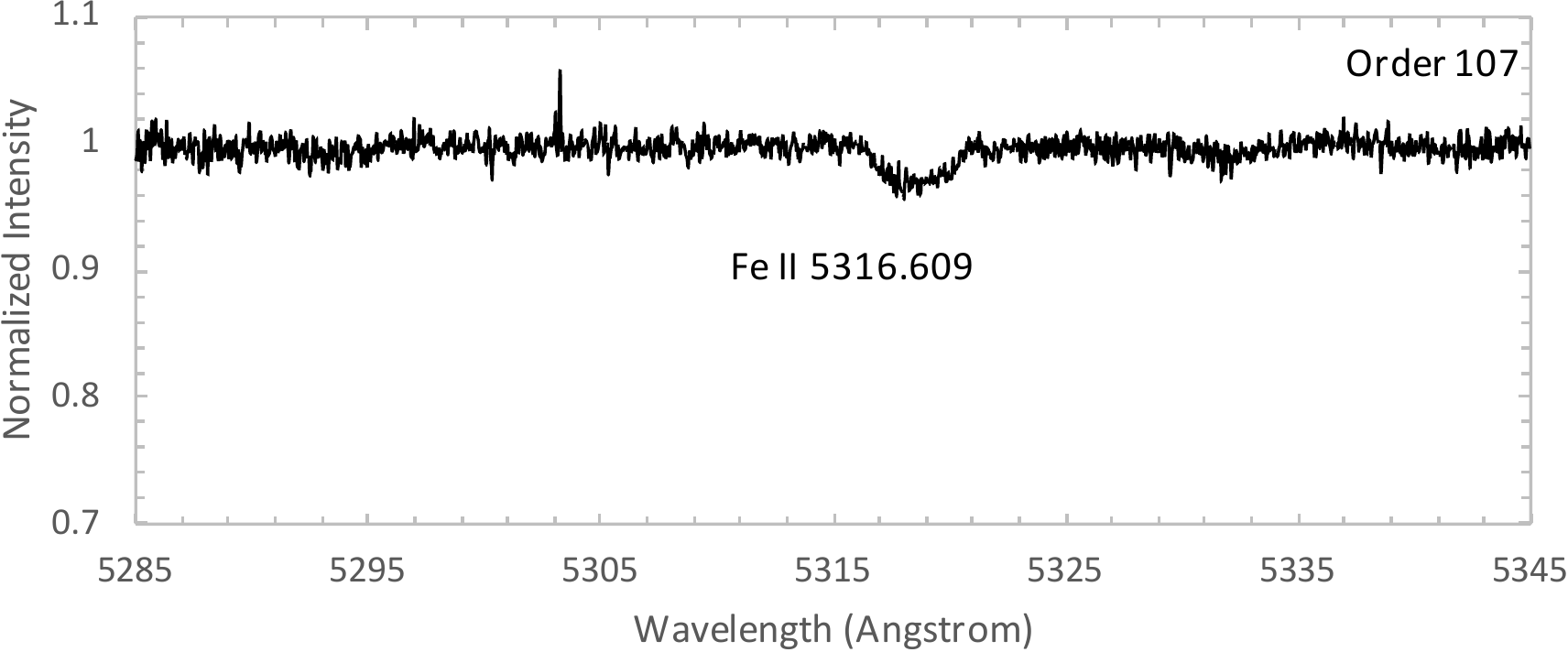} \\
		\includegraphics[scale=0.41]{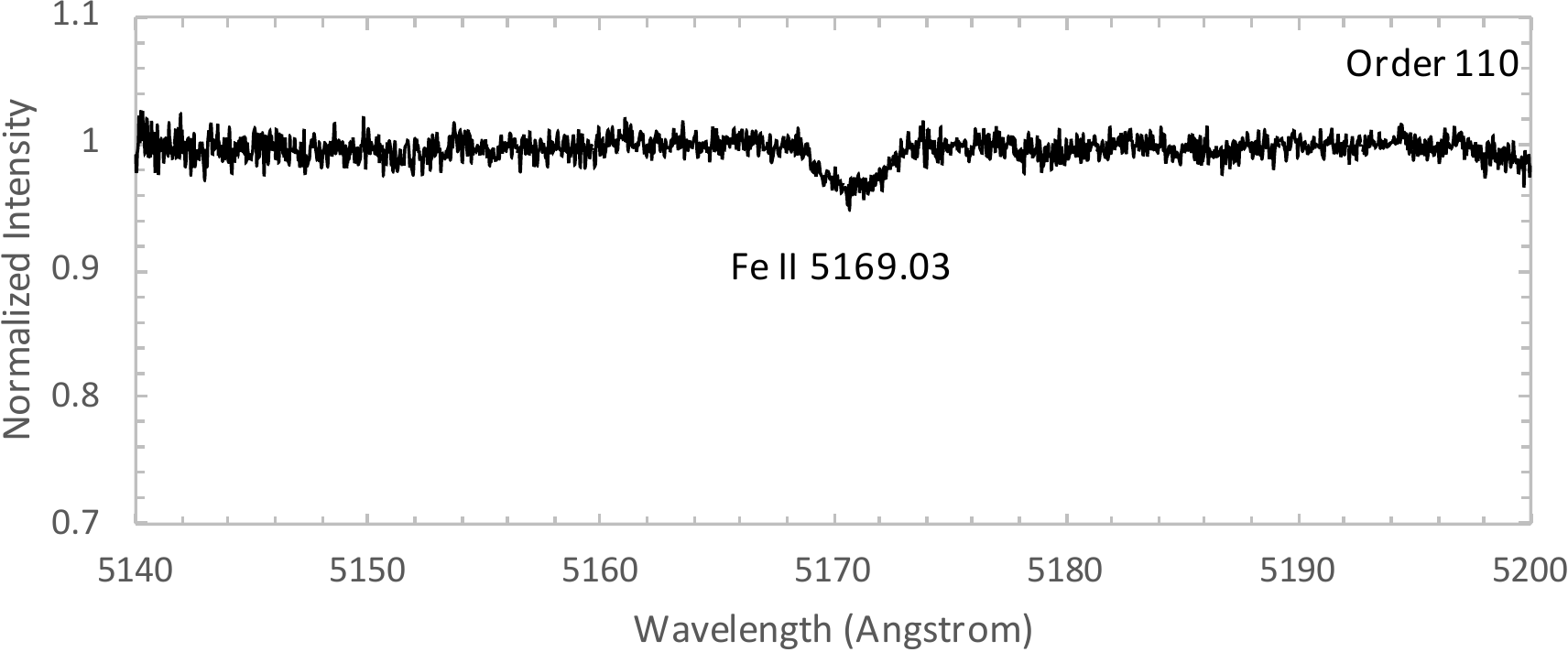}
		\includegraphics[scale=0.41]{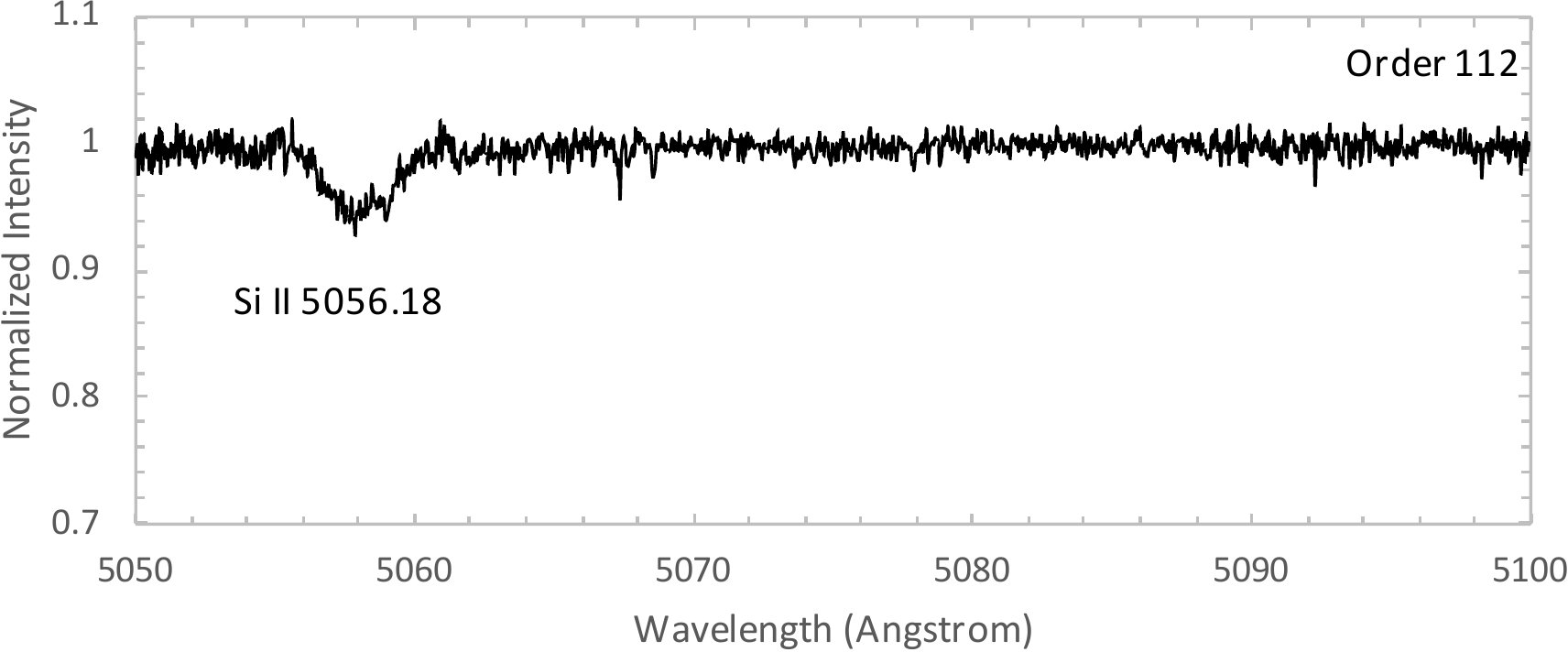} \\
	\includegraphics[scale=0.41]{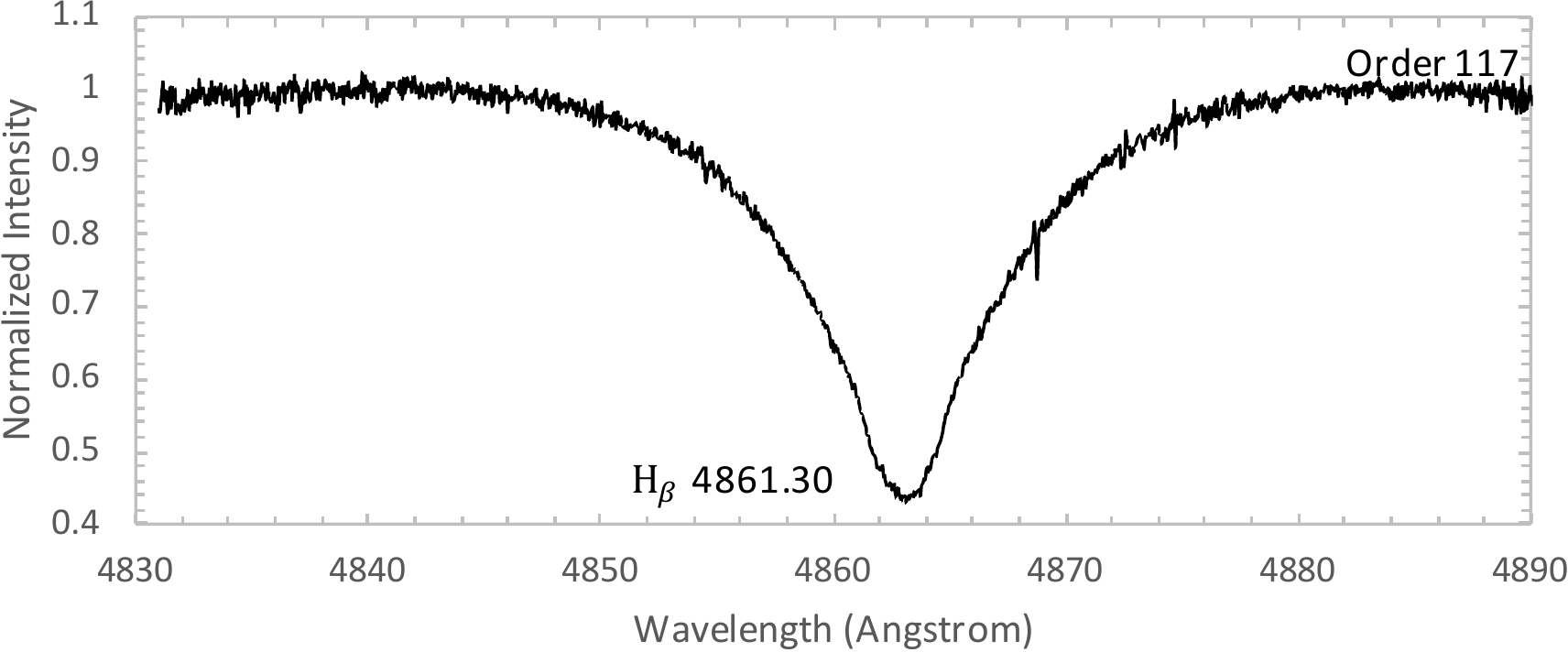}
	\includegraphics[scale=0.41]{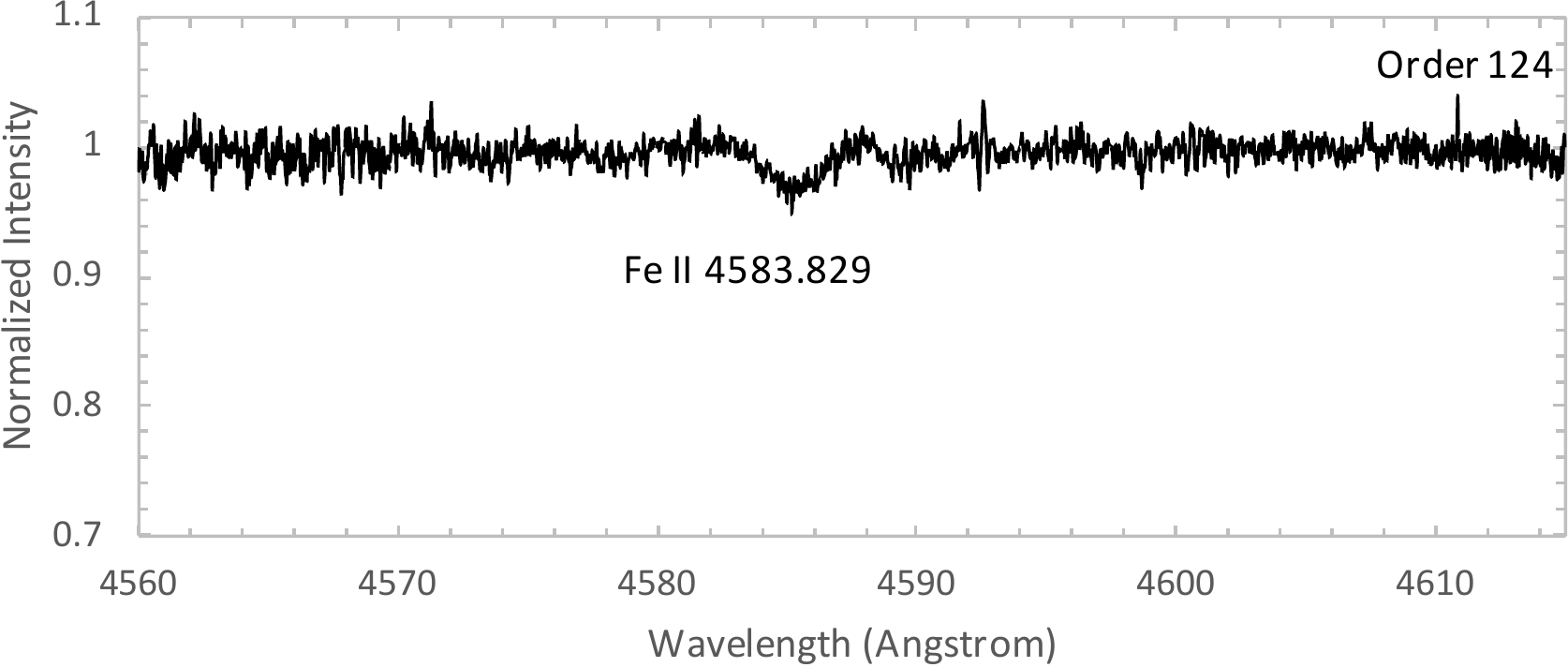} \\
	    \includegraphics[scale=0.41]{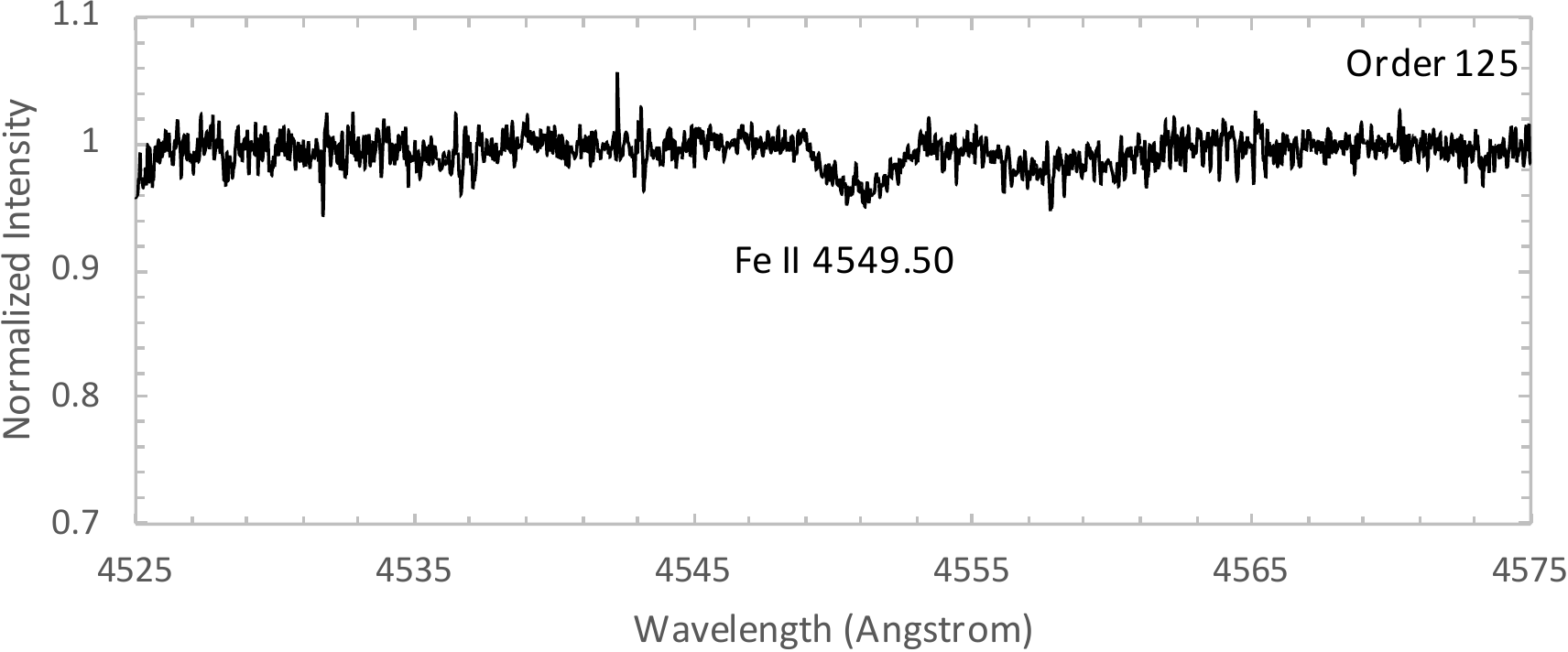}
\end{center}
{Figure B1: Example Plots of Spectral Orders chosen for line identifications of PU Pup. The \'{e}chelle spectrum was taken on the night of December 03, 2015 and at orbital phase of 0.54.
The lines used in RV measurements of PU Pup are indicated in each diagram.
However, as indicated in Section 3.1 and Table 2, mostly strong lines of He I and Si II were used in the RV measurements.
}
\label{orders}
%\end{figure}

% Don't change these lines
%\bsp	% typesetting comment
\label{lastpage}
\end{document}